\batchmode
\makeatletter
\def\input@path{{/home/leoncini/texte/articles/3vortex//}}
\makeatother
\documentclass[10pt]{article}
\usepackage[T1]{fontenc}
\usepackage{graphics}
\usepackage{setspace}

\makeatletter

\providecommand{\LyX}{L\kern-.1667em\lower.25em\hbox{Y}\kern-.125emX\@}
\let\SF@@footnote\footnote
\def\footnote{\ifx\protect\@typeset@protect
    \expandafter\SF@@footnote
  \else
    \expandafter\SF@gobble@opt
  \fi
}
\expandafter\def\csname SF@gobble@opt \endcsname{\@ifnextchar[
  \SF@gobble@twobracket
  \@gobble
}
\edef\SF@gobble@opt{\noexpand\protect
  \expandafter\noexpand\csname SF@gobble@opt \endcsname}
\def\SF@gobble@twobracket[#1]#2{}

 \newcommand{\lyxaddress}[1]{
   \par {\raggedright #1 
   \vspace{1.4em}
   \noindent\par}
 }

\makeatother

\begin{document}

\title{Motion of Three Vortices near Collapse }

\author{X. Leoncini\protect\( ^{1}\protect \), L. Kuznetsov\protect\( ^{2}\protect \)\thanks{
Current adress: Division of Applied Mathematics, Brown University, Providence,
RI 02912
}, and G. M. Zaslavsky\protect\( ^{1,2}\protect \)}

\maketitle

\lyxaddress{\centering {\small \protect\( ^{1}\protect \)Courant Institute of Mathematical
Sciences, New York University, 251 Mercer St., New York, NY 10012, USA }\small }

\lyxaddress{\centering {\small \protect\( ^{2}\protect \)Department of Physics, New York
University, 2-4 Washington Place, New York, NY 10003, USA }\small }

\begin{abstract}
A system of three point vortices in an unbounded plane has a special family
of self-similarly contracting or expanding solutions: during the motion, vortex
triangle remains similar to the original one, while its area decreases (grows)
at a constant rate. A contracting configuration brings three vortices to a single
point in a finite time; this phenomenon known as vortex collapse is of principal
importance for many-vortex systems. Dynamics of close-to-collapse vortex configurations
depends on the way the collapse conditions are violated. Using an effective
potential representation, a detailed quantitative analysis of all the different
types of near-collapse dynamics is performed when two of the vortices are identical.
We discuss time and length scales, emerging in the problem, and their behavior
as the initial vortex triangle is approaching to an exact collapse configuration.
Different types of critical behaviors, such as logarithmic or power-law divergences
are exhibited, which emphasizes the importance of the way the collapse is approached.
Period asymptotics for all singular cases are presented as functions of the
initial vortices configurations. Special features of passive particle mixing
by a near-collapse flows are illustrated numerically. 
\end{abstract}

\section{Introduction}

The importance of point vortices in applications is due to the dominant role
of coherent vortical structures in many \( 2D \) turbulent flows. Experimental
and numerical studies performed in \cite{Tabeling98}-\cite{Carnevale91} have
demonstrated, that in cases of driven or freely decaying \( 2D \) turbulence
a number of concentrated vortices develops out of an originally unstructured
flow. In many situations, dynamics of this finite-size vortices can be reasonably
well approximated by point-vortex models \cite{Zabusky82}-\cite{VFuentes96},
allowing the reduction of the study of a time dependent field, to a simpler
Hamiltonian system of \( N \) interacting particles. Point-vortex dynamics
is an essential ingredient of punctuated Hamiltonian models, which have been
successfully used to describe the evolution of \( 2D \) turbulence after the
emergence of the vortices \cite{Carnevale91,Benzi92,Weiss99}. In these models,
the advection of well-separated vortices is approximated via Hamiltonian point-vortex
dynamics; to account for the change in the vortex population toward smaller
number of bigger vortices, dissipative merging processes \cite{Melander88,Yasuda97}
are included for vortices which have approached each other closer than a certain
critical distance. Point-vortex dynamics is responsible for bringing vortices
together, i.e. it determines what kind of merger processes will occur and how
often.

The number of vortices in a flow can be quite large, in which case a complete
dynamical description gets intractable, but still can yield some important statistical
quantities \cite{Weiss98,Min96}. On the other hand, few-vortex systems can
be investigated in much more detail. Recent interest in this area was mainly
directed towards the study of Lagrangian chaos in different settings \cite{Melezhko92}-\cite{Boffetta96}.
Apart from this, low-dimensional vortex dynamics is essential for the understanding
of the evolution of many-vortex flows, since it describes their \char`\"{}elementary
interactions\char`\"{} \cite{Aref79}. On an unbounded plane one needs at least
three vortices to get a non-trivial dynamics, leading to non-constant inter-vortex
distances. On the other hand, motion of three vortices on an unbounded plane
is always integrable, whereas a generic four-vortex system is already chaotic
\cite{Novikov78,Aref80,Ziglin80}. Hence, a three-vortex system is the only
generic nontrivial case, which allows a complete dynamical analysis. General
classification of different types of three vortex motion, as well as studies
of special cases of particular interest were repeatedly addressed by many authors
\cite{Synge49,Novikov75,Aref79,Aref80,Tavantzis88,Novikov79,Kimura90,Melezhko92},
see \cite{Aref83} for a review.

If we think about point vortices as an approximation of a finite-size patches
of concentrated vorticity, than we have to be sure that they always stay separated
by a distance several times larger than a characteristic size of the patch,
in order for the approximation to be valid (see \cite{Vobseek97} for a direct
comparison of finite-size patch dynamics with its point vortex counterpart).
A natural question is how close can the vortices approach each other during
their motion? A rather surprising answer is that three vortices can be brought
to a single point in finite time by their mutual interaction. Aref \cite{Aref83}
points out that this result was already known to Gr\"{o}bli more than a century
ago. This phenomenon, known as \textit{point-vortex collapse} was studied in
\cite{Aref79,Novikov79,Tavantzis88,Kimura90}. The dynamics of exact collapse
is relatively simple due to a self-similarity of vortex motion in this case:
a triangle formed by vortices (vortex triangle) rotates and shrinks, but stays
similar to the initial one. Vortex triangle area decreases linearly with time,
at a rate determined by vortex strengths and triangle shape \cite{Aref79}.
A spatial reflection of a vortex triangle is equivalent to a time reversal,
so that a reflection of a collapse configuration will experience an infinite
self-similar expansion with the same rate of change of triangle area. The finite-time
collapse of three point vortices into one (or its reverse the splitting of one
vortex into three) changes the topology of the flow field, this feature itself
has important consequences: this phenomenon may play an important role in topologically
driven transitions where vortices are involved, and it conceptually allows the
concentration (collapse) or spreading (splitting) of vorticity even for non-viscous
media.

As a matter of fact, analysis of rare gas of vortex patches, performed in \cite{Zabusky96}
indicates, that in the limit of low vortex density (vortex occupation less than
one-two percent), strong interactions occur only between three vortices in a
specific way, similar to the \textit{three point-vortex collapse}. These processes
may be thought of as resonance interactions in the vortex gas.

For the collapse to happen certain conditions on vortex strengths and initial
positions have to be satisfied exactly, so that in a real system, probability
of this event is zero: true resonances do not occur. However, actual vortices
have some characteristic size, and when distance between them gets comparable
with their size, vortices experience a considerable distortion, and may merge
together, which means, that from a broader physical point of view, a collapse
can be thought of as a process which brings vortices close enough (up to their
size) to each other. For three initially well-separated vortices to collapse
in the above sense, their strength and positions may satisfy \char`\"{}resonance
conditions\char`\"{} only approximately. We refer to this kind of motion as
near-collapse dynamics. Its importance stems from the fact, that inter-vortex
distances are changed considerably (by orders of magnitude) during the motion,
which brings the system to a different length-scale, where new physical mechanisms
enter into play.

The main goal of this article is to describe different scenarios of the 3-vortex
system in near-collapse dynamics, in other words we perform a quantitative analysis
of the motion in near-collapse situations, when the resonance initial conditions
are slightly distorted, and/or the vortex strengths are not precisely tuned.
These results are briefly resumed in tables \ref{Table1}-\ref{Table3} included
in Section 3. We restrict our attention to the case when two of the vortices
are identical, for which it is possible to map the dynamics of the vortices
to a motion of a particle in a one dimensional potential. Our description is
tailored to obtain the dynamical characteristics of the motion, such as the
intervals of vortex distance variation, period of relative motion, etc, and
even in a simplified situation of two equal-strength vortices, we have found
a rich pattern of vortices dynamics. Shortly, depending on the parameters of
the system, the vortices can perform periodic dynamics with different singular
periods as well as periodic dynamic with non-singular period. More importantly
we have found that the singularity of the periods can be of the logarithmic
or pole type. This information about possible time scalings of the near collapse
motion should be taken into account when many-vortex systems are considered
and, particularly it is significant for studying advection in 3-vortex systems.
We will come to this issue in a forthcoming publication. 

In Section 2 the basic equations of vortex motion, their symmetries and corresponding
first integrals are specified. Effective Hamiltonian for a three-vortex system
with two identical vortices is introduced in Section 3. Using this Hamiltonian,
different types of near-collapse dynamics are classified and analyzed. In Section
4 we discuss various routes to collapse, i.e. the manner in which near-collapse
motion types approach self-similarly contracting (expanding) solutions as initial
configuration is taken closer to the exact resonance. In Section 5 we give an
illustration of how the approach to collapse affects passive particle advection.

\section{Dynamical equations}

Point vortices are exact solutions of Euler equation (see for example \cite{Machioro94});
which writes for the vorticity \( \Omega  \)
\begin{equation}
\label{vorti}
\frac{\partial \Omega }{\partial t}+\left[ \Omega ,\psi \right] =0\: ,
\end{equation}
 where \( [\cdot ,\cdot ] \) is the usual Poisson bracket, and \( \psi  \)
is the stream function; for two dimensional motion of incompressible fluid \( \Omega =-\nabla ^{2}\psi  \).
Equation (\ref{vorti}) expresses the conservation of generalized vorticity
along the path lines of the flow.

A point vortex system is defined by a vorticity distribution given by a superposition
of Dirac functions: 
\begin{equation}
\Omega ({\textbf {R}},t)=\sum _{i=1}^{N}k_{i}\delta \left( {\textbf {x}}-{\textbf {x}}_{i}(t)\right) \: ,
\end{equation}
 where \( {\textbf {x}} \) is a vector in the plane of the flow, \( k_{i} \)
is the circulation of \( i \)-th vortex, \( N \) is the total number of vortices,
and \( {\textbf {x}}_{i}(t) \) is the vortex position at time \( t \). Using
this expression for the vorticity, and solving Poisson equation, in the Euler
case, one obtains a stream function of a point vortex system. By Helmholtz theorem
\cite{Saffman95}, the motion of vortices is determined by the value of the
velocity field at the position of the vortex. The equation of a point vortex
motion is: 
\begin{equation}
\label{pvor+}
k_{i}\frac{d{\textbf {x}}_{i}}{dt}={\textbf {e}}_{z}\wedge \frac{\partial \psi }{\partial {\textbf {x}}_{i}}\: .
\end{equation}
 where \( {\textbf {e}}_{z} \) is the unit vector perpendicular to the flow.
The motion has an Hamiltonian structure with the two Cartesian coordinates as
conjugated variables, and by convention we shall drop \( \psi  \) and use the
notation \( H \) instead. The Hamiltonian \( H \) is given by: 
\begin{equation}
\label{ham}
H=\frac{1}{2\pi }\sum _{i>j}k_{i}k_{j}U(x_{ij})
\end{equation}
 where \( x_{ij}=|{\textbf {x}}_{i}-{\textbf {x}}_{j}| \) and the interaction
potential \( U(x)=-\log (x) \) in case of an unbounded plane. The Hamiltonian
(\ref{ham}) is invariant under translation and rotation, which implies both
the conservation of the vortex momentum: 
\begin{equation}
\label{momen1}
{\textbf {P}}\equiv \sum ^{N}_{i=1}k_{i}{\textbf {x}}_{i}(t)=const.
\end{equation}
 and vortex angular momentum \( L{\textbf {e}}_{z} \) with: 
\begin{equation}
\label{momen2}
L^{2}\equiv \sum _{i=1}^{N}k_{i}{\textbf {x}}_{i}^{2}(t)=const.
\end{equation}

\begin{figure}[!h]
\par\centering \resizebox*{8.5cm}{!}{\includegraphics{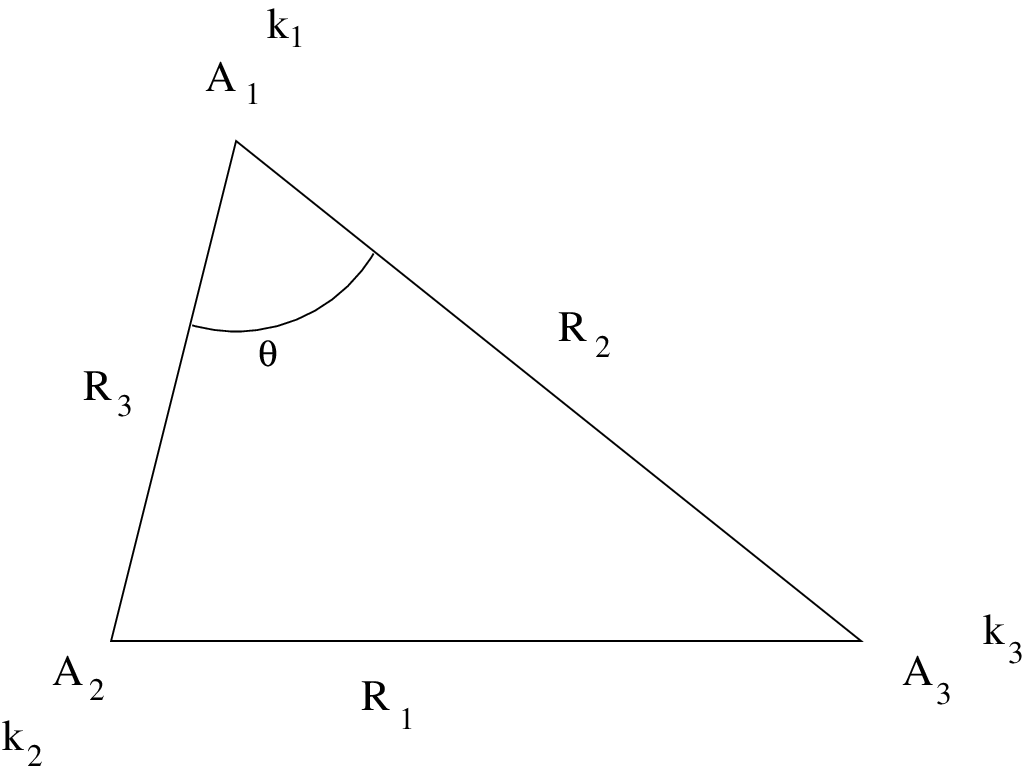}}  \par{}

\caption{Notation chosen for the problem\label{NotConvfig}}
\end{figure}

In this paper we consider a case of three vortices in an unbounded domain. We
restrict our attention to the relative motion of vortices, taking inter-vortex
distances \( R_{i} \) as prime variables. The notation we use is illustrated
in Fig.~\ref{NotConvfig}. The invariance of the Hamiltonian (\ref{ham}) under
translations allows us a free choice of the coordinate origin, which we put
to the center of vorticity (when it exists). Then, the other two constants of
motion (energy and angular momentum) written in a frame independent form become:

\begin{equation}
\label{constantmotion1}
\left\{ \begin{array}{l}
H=-\frac{1}{2\pi }\left[ k_{1}k_{2}\ln R_{3}+k_{1}k_{3}\ln R_{2}+k_{3}k_{2}\ln R_{1}\right] \\
K=\left[ \left( \sum _{i}k_{i}\right) L^{2}-P^{2}\right] =k_{1}k_{2}R^{2}_{3}+k_{1}k_{3}R^{2}_{2}+k_{3}k_{2}R^{2}_{1}\: .
\end{array}\right. 
\end{equation}
 The equations of motion (\ref{pvor+}) yield the following non canonical system
for the dynamics of the inter-vortex distances:

\begin{equation}
\label{equmotion1}
\left\{ \begin{array}{l}
k^{-1}_{1}R_{1}\dot{R}_{1}=A/\pi (R^{-2}_{2}-R^{-2}_{3})\\
k^{-1}_{2}R_{2}\dot{R}_{2}=A/\pi (R^{-2}_{3}-R^{-2}_{1})\\
k^{-1}_{3}R_{3}\dot{R}_{3}=A/\pi (R^{-2}_{1}-R^{-2}_{2})\: ,
\end{array}\right. 
\end{equation}
 where \( A \) is the area of the triangle \( A_{1}A_{2}A_{3} \) (see Fig.
1), and \( \dot{x} \) refers to the time derivative of \( x \), (See \cite{Aref79,Tavantzis88}
for details).

The physical system being defined, we shall now focus on the near collapse configuration.
As mentioned earlier, for the three point vortex problem, under certain conditions,
which depends both on the initial conditions, and the vortex strengths, the
motion is self similar, leading either to the collapse of the three vortices
in a finite time, or by time reversal, to an infinite expansion of the triangle
formed by the vortices. These conditions of a collapse or an infinite expansion
of the three point vortices are easily found. The first condition is immediate
as for a collapse to occur \( K \) has to vanish. For the second condition
we just need to look for conditions to obtain a scale invariant Hamiltonian.
With such spirit let us divide all lengths \( R_{i} \) by a common factor \( \lambda  \)
in the Hamiltonian \( H \). We readily obtain \( H'(\lambda )=H+\left( \sum k_{i}k_{j}\right) \ln \lambda  \),
we then obtain the collapse conditions:
\begin{equation}
\label{geomconditions}
K=0
\end{equation}
\begin{equation}
\label{strengconditions}
\sum _{i}\frac{1}{k_{i}}=0\: ,
\end{equation}
 \emph{i.e}, the harmonic mean of the vortex strengths (\ref{strengconditions})
and the total angular momentum in its frame free form (\ref{geomconditions}),
are both zero. Near a collapse configuration, the two conditions (\ref{geomconditions}),
(\ref{strengconditions}) allow two different ways to approach the singularity.
Namely, we can change initial conditions which changes the value of \( K \),
or change the vortex strength and modify the harmonic mean. Unfortunately the
motion of the three vortices even though integrable is not easily computed analytically.
It then seemed useful to restrict the problem to some less general case, assuming
that the basic mechanisms of collapse should not differ much in the general
case. To emphasize this last statement, let us first consider a general collapse
situation, for which the total strength is \( k_{tot} \). The following conditions
are then fulfilled 
\begin{equation}
\label{justif1}
\sum _{i}k_{i}=k_{tot}\: ,\sum _{i}\frac{1}{k_{i}}=0\: ,
\end{equation}
 which are equivalent to 
\begin{equation}
\label{justif2}
\sum _{i}k_{i}=k_{tot}\: ,\sum _{i}k^{2}_{i}={k_{tot}}^{2}\: .
\end{equation}
 The values for the vortex strengths corresponding to a collapse configuration
are therefore located on the circle resulting from the intersection of the sphere
of radius \( {k_{tot}}^{2} \) and the plane imposing a total vortex strength
equal to \( k_{tot} \). Since we have a circle, we shall assume that, from
this point of view, no special configuration exists, and any should be sufficient
to describe qualitatively the different possible behaviors.

In the following we will be considering the case, when two of the three vortices
are identical. Since the rescaling of time scales leads to a freedom in vortex
strength normalization, we will assume 
\begin{equation}
k_{2}=k_{3}=1
\end{equation}
 Then, in order for a collapse to happen, the strength of the third vortex has
to satisfy (\ref{strengconditions}) which gives the resonance value of the
first vortex: \( k_{1c}=-1/2 \). We will be mostly interested in the situations,
when \( k_{1} \) is negative and close to its resonance value, so we will denote
\begin{equation}
k\equiv |k_{1}|\: ,
\end{equation}
 and define the deviation from the resonance as 
\begin{equation}
\delta =1/2-k\: .
\end{equation}

\section{Near-collapse dynamics of vortices \label{section2}}

In the case when two out of three vortices are identical, there exists a convenient
representation of the system dynamics in terms of a motion of a particle in
a one-dimensional potential, parametrically depending on vortex initial condition.
To derive this representation, we introduce a new set of variables: \( X=R^{2}_{1} \),
\( Y=R^{2}_{2}R^{2}_{3} \), and \( Z=R^{2}_{2}+R^{2}_{3} \). The constants
of motion (\ref{constantmotion1}) can be written as: 
\begin{equation}
\label{constantmotion2}
\left\{ \begin{array}{l}
\Lambda =e^{4\pi H}=Y^{k}/X\\
K=X-kZ\: .
\end{array}\right. 
\end{equation}
 where new parameter \( \Lambda  \) is introduced instead of \( H \) in order
to simplify formulas.

An equation on \( X \) can be directly obtained from the first equation of
(\ref{equmotion1}), squaring it gives: 
\begin{equation}
\label{firststepsquare}
\dot{X}^{2}=\frac{4}{\pi ^{2}}A^{2}k^{2}\frac{(R_{2}^{2}+R^{2}_{3})^{2}-4R^{2}_{2}R^{2}_{3}}{R_{2}^{4}R^{4}_{3}}=\frac{4}{\pi ^{2}}A^{2}k^{2}\frac{Z^{2}-4Y}{Y^{2}}\: .
\end{equation}
 Square of the vortex triangle area \( A \) can be found from geometrical identities
\begin{equation}
\label{areacomp}
\left\{ \begin{array}{l}
A=\sqrt{Y}|\sin \theta |/2\\
X=Z-2\sqrt{Y}\cos \theta \: ,
\end{array}\right. 
\end{equation}
 which leads to 
\begin{equation}
\label{areacomp2}
16A^{2}=4Y-(X-Z)^{2}\: .
\end{equation}
 Using the expression for the constants of motion (\ref{constantmotion2}),
we obtain:

\begin{equation}
\label{equmotion2}
\dot{X}^{2}=\frac{1}{4\pi ^{2}}\frac{[4k^{2}Y-(K-(1-k)X)^{2}][(K-X)^{2}-4k^{2}Y]}{k^{2}Y^{2}},
\end{equation}
 where \( Y=(\Lambda X)^{1/k} \). This equation has a form of an energy conservation
law for a particle of mass \( 1 \) and zero total energy moving in a potential
\( V(X;\Lambda ,K,k) \), defined by 
\begin{equation}
\label{potential}
V(X;\Lambda ,K,k)\equiv \frac{{[(K-(1-k)X)^{2}-4k^{2}Y][(X-K)^{2}-4k^{2}Y]\}}}{8\pi ^{2}k^{2}Y^{2}}\: .
\end{equation}
 Indeed, equation (\ref{equmotion2}) can be rewritten as 
\begin{equation}
\label{Hamilteff}
H_{eff}(\dot{X},X;\Lambda ,K,k)\equiv P^{2}/2+V(X;\Lambda \, K,k)=0\: ,
\end{equation}
 with Hamiltonian equations 
\begin{equation}
\dot{X}=\partial H_{eff}/\partial P\equiv P\: ,\hspace {10mm}\dot{P}=-\partial H_{eff}/\partial X\: .
\end{equation}
 Thus, dynamics of vortex configuration is governed by an effective Hamiltonian
\( H_{eff} \), where the shape of the potential well \( V \) depends on the
initial vortex positions through the values of first integrals \( \Lambda  \)
and \( K \) (\ref{constantmotion2}), and the strength of the first vortex
\( k \).

Effective Hamiltonian representation (\ref{Hamilteff}), gives a number of advantages,
allowing to use simple standard techniques for one-dimensional conservative
systems to find dynamical properties of vortex motion. In a sense, the problem
is reduced to determining the shape of the potential \( V \) as a function
of its parameters.

Before proceeding to the analysis of the types of potentials, emerging in our
problem, we have to mention the restrictions, imposed on the system by the fact
that three vortices form a triangle. Since this was not taken into account in
the derivation of the effective Hamiltonian (\ref{Hamilteff}), it may put additional
boundaries on the possible values of \( X \). Note, that since the effective
energy \( H_{eff} \) in (\ref{Hamilteff}) is always zero, a motion exists
only on the segments where the potential is negative. Therefore, triangle inequalities
and the condition \( V\leq 0 \) define the physical domain of \( X \). Triangle
inequalities 
\begin{equation}
\label{trian1}
|R_{2}-R_{3}|\leq R_{1}\leq R_{2}+R_{3}\: ,
\end{equation}
 written in terms of new variables can be obtained by taking the square of (\ref{trian1}):
\( Z-2\sqrt{Y}\leq X\leq Z+2\sqrt{Y} \), which is equivalent to 
\begin{equation}
\label{trianglecondition}
(X(k-1)+K)^{2}-4k^{2}Y\leq 0\: .
\end{equation}
 The condition (\ref{trianglecondition}) is equivalent to the positiveness
of the right hand side of equation (\ref{areacomp2}) and translates that the
square of the area of the triangle \( A_{1}A_{2}A_{3} \) is positive. The condition
\( V(X)\leq 0 \) together with the triangle inequalities (\ref{trianglecondition})
imposes then 
\begin{equation}
\label{condition2}
(K-X)^{2}-4k^{2}Y\geq 0\: ,
\end{equation}
 Recalling that \( Z>0 \), we have \( X\geq K \), which together with inequality
(\ref{condition2}), gives finally 
\begin{equation}
\label{finalcondition}
X\geq K+2k\sqrt{Y}\: .
\end{equation}

Now we have a precise information on where the motion should lie, and can proceed
to the study of different regimes of near-collapse dynamics. Although the effective
potential representation is valid for any \( k\neq 0 \), we will restrict our
study (and our definition of ``near-collapse'') to the interval \( k\in ]0;1[ \),
since the special cases \( k=1 \) (escaping vortex pair) and \( k=2 \) (neutral
tripole), as well as the degenerate case \( k=0 \) (two vortices and a passive
particle) bring up their specific singularities, unrelated to the collapse phenomenon.
Let us consider different regimes such that each regime corresponds to a class
of qualitatively similar motions \cite{Aref79}. Using the effective Hamiltonian
representation (\ref{Hamilteff}), we infer that different regimes may be encountered
corresponding to the number of roots of the potential \( V(X;\Lambda ,K,k) \)
lying within the physical domain. This number changes with variation of the
parameters \( (\Lambda ,K,k) \).

Note, that the absolute value of \( K \) depends on the length units and can
be scaled out of the problem. Indeed, the potential \( V \) and the physical
domain boundary (\ref{finalcondition}) are invariant under the scaling transformation
(compare to (\ref{constantmotion2})) 
\begin{equation}
\label{rescaling1}
X\rightarrow \frac{X}{|K|},\hspace {1cm}\Lambda \rightarrow \Lambda |K|^{\delta }\: ,
\end{equation}
 and the problem can be reduced to the study of the three following cases 
\begin{equation}
\label{Threecases}
K=1\: ,\; K=-1\: ,\; K=0\; ,
\end{equation}
 where the last case is singular. The different situations encountered have
been condensed in the tables \ref{Table1}-\ref{Table3}, which are presented
at the end of this Section. This is intended to briefly summarize the result
of this work and allow the reader to browse through this paper more easily.

In the following we will discuss in details the different types of motions and
critical situation encountered; most of the figures illustrating the problem
will correspond to the values (\ref{Threecases}), although we will keep \( K \)
as a scale parameter in formulas, keeping in mind, that only its sign is relevant
to distinguish different regimes. We insist that the scaling of \( \Lambda  \)
in (\ref{rescaling1}), which can be rewritten in terms of vortex energy (\ref{constantmotion1})
as 
\begin{equation}
\label{Hrescaling}
H\rightarrow H+\delta \ln |K|\: ,
\end{equation}
 leaves \( \Lambda  \) and \( H \) unchanged, when \( \delta =0 \), i.e.
when vortex strengths exactly satisfy the collapse condition (\ref{strengconditions}).
The scales of the motion are determined by the value of the angular momentum
\( K \), while from a pure energetic point of view the motion would be thought
as scale invariant.

\subsection{Critical situations for \protect\protect\protect\( K\ne 0\protect \protect \protect \)\label{criticsitu_par}}

To detect bifurcations, leading to the appearance of new roots of the potential,
we look for the degenerate roots of (\ref{potential}). As a first step, we
just want to find these roots, leaving their detailed interpretation for the
following paragraphs. Such roots \( X_{c} \) exist only for certain critical
values \( \Lambda _{c} \) of the energy parameter \( \Lambda  \), and can
be found from: 
\begin{equation}
\label{criticalconditions}
\left\{ \begin{array}{l}
V(X_{c},\Lambda _{c})=0\\
\partial V/\partial X(X_{c},\Lambda _{c})=0\: 
\end{array}\right. 
\end{equation}
 The above system can be easily solved if one notices, that the potential (\ref{potential})
is written in a factorized form: \( V=V_{1}*V_{2}/(8\pi ^{2}k^{2}Y^{2}) \),
with 
\begin{equation}
V_{1}\equiv [(K-(1-k)X)^{2}-4k^{2}Y],\hspace {1cm}V_{2}\equiv [(K-X)^{2}-4k^{2}Y]
\end{equation}
 Then, if a solution \( X_{c},\, \, \Lambda _{c} \) exists and is different
from \( X=0 \), it should be either a double zero of \( V_{1} \) or \( V_{2} \),
or a zero of both \( V_{1} \) and \( V_{2} \).

We start from finding a double zero of \( V_{2} \), which yields a system similar
to (\ref{criticalconditions}), where \( V \) is substituted by \( V_{2} \),
that leads to: 
\begin{equation}
\label{criticcond1}
\left\{ \begin{array}{l}
\left( X-K\right) ^{2}-4k^{2}Y=0\\
2kX\left( X-K\right) -4k^{2}Y=0\: 
\end{array}\right. 
\end{equation}
 and after some substitutions we readily obtain, 
\begin{equation}
\label{criticsitu1}
\left\{ \begin{array}{l}
X_{c_{1}}=K/2\delta \\
\Lambda _{c_{1}}=\left( K/2\delta \right) ^{-2\delta }\; 
\end{array}\right. 
\end{equation}
 For \( X_{c_{1}} \) to be positive, \( K \) and \( \delta  \) must have
the same sign, so this bifurcation pertains to the physical region only for
the cases \( K>0,\, \, k<1/2 \) and \( K<0,\, \, k>1/2 \) (it is easy to check
that substitution of (\ref{criticsitu1}) into (\ref{finalcondition}) turns
the latter into an identity). Another solution, \( X=K,\, \, \Lambda =0 \),
corresponds to the situation when the negative vortex merges with one of the
two others, reducing the system to a two-vortex case. Notice that \( X_{c_{1}} \)
diverge as vortex strength collapse condition (\ref{strengconditions}) is approached
(\( \delta \rightarrow 0,k\rightarrow 1/2 \)), which announces a special behavior
of the case \( \delta =0 \), requiring a special treatment.

Double zeros of \( V_{1} \) are found in the same manner. After some algebra
we obtain: 
\begin{equation}
\label{criticsitu2}
\left\{ \begin{array}{l}
X_{c_{2}}=X_{c_{1}}/\left( 1-k\right) \\
\Lambda _{c_{2}}=(1+2\delta )\Lambda _{c_{1}}/2\; 
\end{array}\right. 
\end{equation}
 and a solution \( X=K/(1-k),\, \, \Lambda =0 \) for the merged situation.
Being proportional to \( X_{c_{1}} \), \( X_{c_{2}} \) diverges in the same
manner as \( \delta \rightarrow 0 \), and requires sign of \( K \) and \( \delta  \)
to be the same, in order to lie in a physical range. It also diverges when \( k\rightarrow 1 \),
as we have mentioned earlier, this is the case of a scattering of a neutral
vortex pair on a vortex, which is out of the scope of the present paper. Finally
we note, that since \( V_{1} \) is proportional to the area of the vortex triangle
\( A \), the critical equilibrium position \( X_{c_{2}} \) corresponds to
an aligned vortex configuration; in a similar manner we deduce that \( X_{c_{1}} \)
corresponds to an equilateral configuration \cite{Aref79,Tavantzis88}.

A third possibility, when both \( V_{1}=0 \) and \( V_{2}=0 \), leads to the
following system: 
\begin{equation}
\label{criticcond3}
\left\{ \begin{array}{l}
\left( X-K\right) ^{2}-4k^{2}Y=0\\
(K-(1-k)X)^{2}-4k^{2}Y=0
\end{array}\right. 
\end{equation}
 which yields: 
\begin{equation}
\label{criticsitu3}
\left\{ \begin{array}{l}
X_{c_{3}}=2K/(2-k)\\
\Lambda _{c_{3}}=\left[ K/\left( 3+2\delta \right) \right] ^{-2\delta }/4\; .
\end{array}\right. 
\end{equation}
 This critical case also corresponds to an aligned configuration, where the
negative vortex of strength \( -k \) is in the middle between the two identical
vortices. In near collapse situation \( 2-k>0 \), and since \( X_{c_{3}} \)
has to be positive, this critical situation appears only for positive total
vortex angular momentum \( K>0 \). Here a merged solution also exists (\( X=0,\, \, \Lambda =+\infty  \)),
which corresponds to the merging of the two identical vortices. We also notice
that the divergence of \( X_{c_{3}} \) for \( k=2 \) corresponds to a special
case of a neutral tripole, when the total vorticity is zero, and the center
of vorticity is not defined.

Now we turn our attention to the three different specific cases distinguished
by the sign of \( K \). For each case, we will indicate which of the above
found bifurcations do occur, and study the phase space trajectories of the effective
Hamiltonian (\ref{Hamilteff}) for different regimes.

\subsection{The case \protect\protect\protect\( K>0\protect \protect \protect \), \protect\protect\( k\neq 1/2\protect \protect \)
\label{K>0para}}

\begin{figure}[!h]
{\par\centering \resizebox*{8.5cm}{!}{\includegraphics{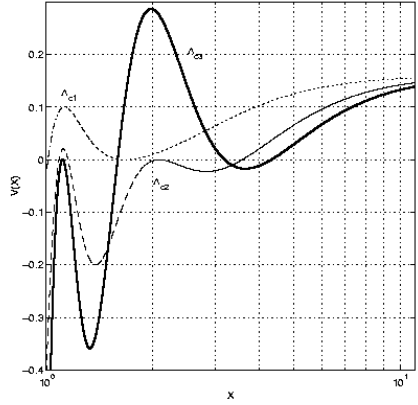}} \par}

\caption{The potential \protect\( V(X)\protect \) for the three critical cases. \protect\( k=0.2\protect \).
For \protect\( \Lambda =\Lambda _{c_{1}}\protect \) the motion appears, for
\protect\( \Lambda =\Lambda _{c_{2}}\protect \) two motions are possible, and
for \protect\( \Lambda =\Lambda _{c_{3}}\protect \) an unstable aligned configuration
is approached as \protect\( X\rightarrow X^{+}_{c_{3}}\protect \).\label{Fig_3criticpot}}
\end{figure}

From the previous Section, we know that the number of critical energies for
this case depends on the sign of \( \delta  \), i.e. whether \( k>1/2 \) or
\( k<1/2 \). A special case \( k=1/2,\, \, \delta =0 \) will be treated separately.
We start from the case \( k<1/2 \), when all three critical cases (\ref{criticsitu1}),
(\ref{criticsitu2}), (\ref{criticsitu3}) belong to the physical region. If
we keep the value of \( k \) fixed, and vary the energy, we will encounter
various motion regimes, separated by the three critical energies. To illustrate
the critical situations we plotted the potential \( V \) as a function of \( X \)
for the three different critical energies for \( k=0.2 \) in Fig.~\ref{Fig_3criticpot}.
The critical energy \( \Lambda _{c_{1}} \) corresponds to the maximum possible
value for which motion can exist; for \( \Lambda =\Lambda _{c_{2}} \) an unstable
equilibrium (saddle point) appears, and motion becomes aperiodic; and in the
last case of \( \Lambda =\Lambda _{c_{3}} \), another saddle point appears
right on the border of the physical region (corresponding to unstable aligned
configuration).

\begin{figure}[!h]
{\par\centering \resizebox*{8.5cm}{!}{\includegraphics{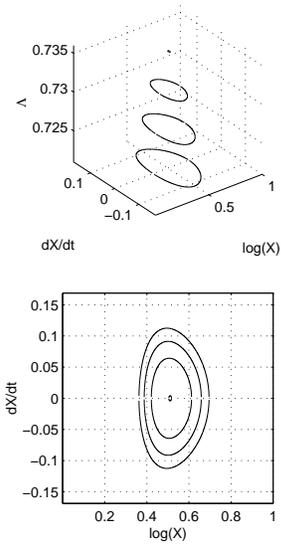}} \par}

\caption{Phase space portrait close to \protect\( \Lambda _{c_{1}}\protect \) for different
energies for the case \protect\( K=1\protect \) and \protect\( k=0.2\protect \).
The bottom figure is the top view of the upper one. Once the first critical
situation is reached a motion is possible.\label{figK>0k<05_1}}
\end{figure}

\begin{figure}[!h]
{\par\centering \resizebox*{8.5cm}{!}{\includegraphics{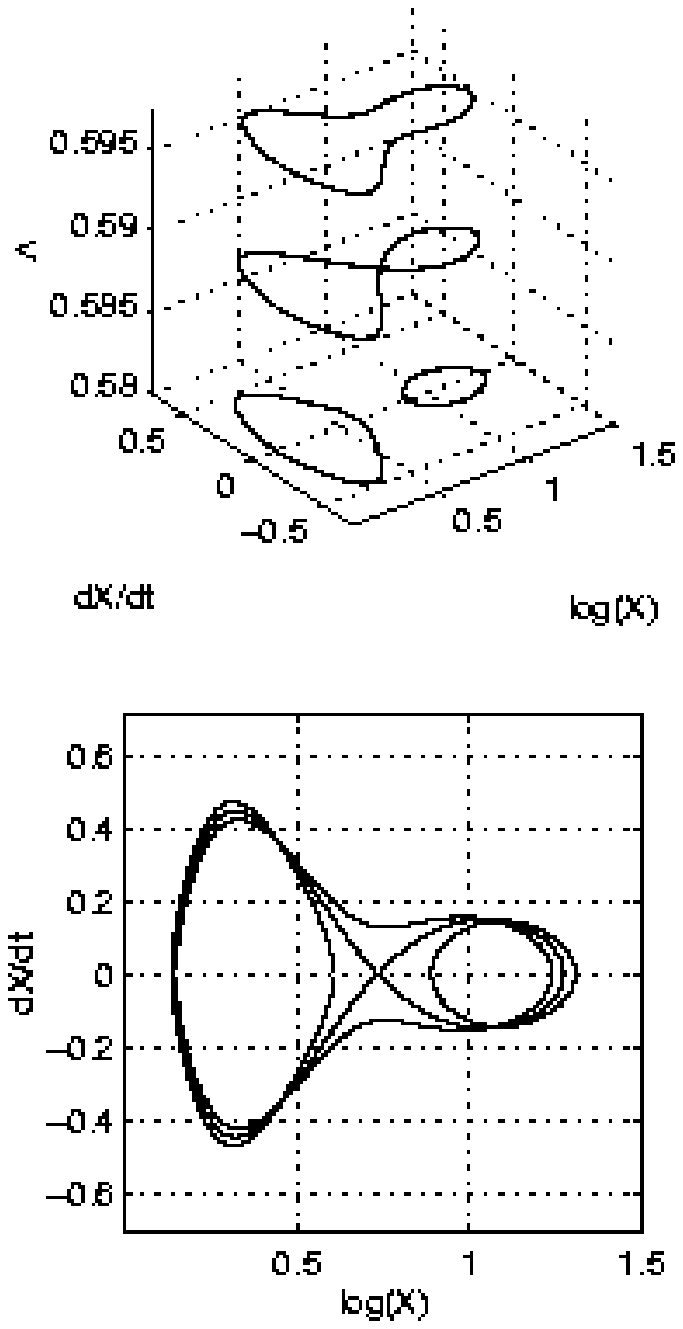}} \par}

\caption{Phase space portrait close to \protect\( \Lambda _{c_{2}}\protect \) for different
energies for the case \protect\( K=1\protect \) and \protect\( k=0.2\protect \).
The bottom figure is the top view of the upper one. Once the second critical
situation is reached, the single possible trajectory reaches the separatrix,
and the splitting in two possible motions occurs.\label{figK>0k<05_2}}
\end{figure}

\begin{figure}[!h]
{\par\centering \resizebox*{8.5cm}{!}{\includegraphics{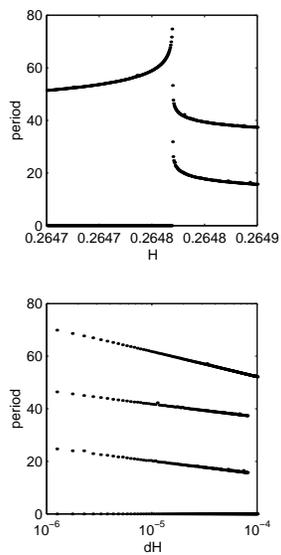}} \par}

\caption{Period of the motion versus energy (\protect\( \Delta H=|H(X)-H_{c_{2}}|\protect \))
for \protect\( k=0.2\protect \) close to \protect\( \Lambda =\Lambda _{c_{2}}\protect \).
We notice the logarithmic behavior of the divergence, we note also that the
sum of the slopes of the two right branches (lower) is equal to the slope of
the left branch (upper).\label{figK>0k<05period}}
\end{figure}

\begin{figure}[!h]
{\par\centering \resizebox*{8.5cm}{!}{\includegraphics{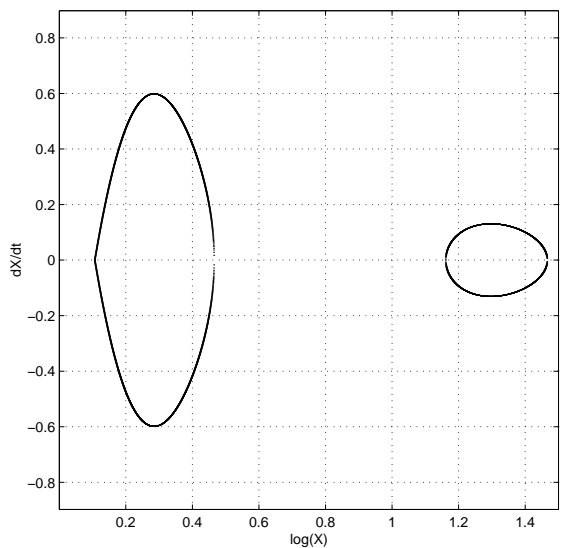}} \par}

\caption{Phase space portrait for \protect\( \Lambda =\Lambda _{c_{3}}\protect \) for
the case \protect\( K=1\protect \) and \protect\( k=0.2\protect \). A splitting
occurs like for \protect\( \Lambda =\Lambda _{c_{2}}\protect \) but the triangle
conditions prevent the existence of a third branch. We notice the sharp angle
at \protect\( X=X_{c_{3}}\protect \). \label{figK>0k<05_3}}
\end{figure}

\begin{figure}[!h]
{\par\centering \resizebox*{8.5cm}{!}{\includegraphics{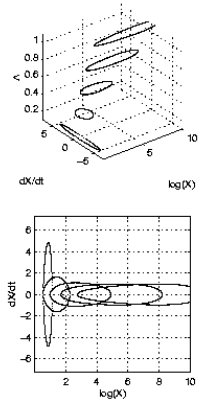}} \par}

\caption{Phase space portrait for different values of \protect\( \Lambda \protect \)
in the case \protect\( K=1\protect \) and \protect\( k=0.55\protect \). The
bottom figure is the top view of the upper one.\label{figK>0k>05} The motion
remains periodic. As \protect\( \Lambda \protect \) increases, we notice the
growth of the range of length scales explored by the motion.}
\end{figure}

Motions regimes for different energy ranges are listed below (See Table \ref{Table1}):

\begin{enumerate}
\item \( \Lambda >\Lambda _{c_{1}} \) Motion impossible. 
\item \( \Lambda _{c_{1}}>\Lambda >\Lambda _{c_{2}} \) The potential has 2 zeros
in the physical region, there exists a single type of periodic motion (see Fig.~\ref{figK>0k<05_1}).
When \( \Lambda \rightarrow \Lambda _{c_{2}} \), the period diverges logarithmically
\begin{equation}
\label{pest1}
T(\Lambda )\sim 1/2\ln |\Lambda -\Lambda _{c_{2}}|\: ,
\end{equation}
 (see Fig.~\ref{figK>0k<05period}) due to proximity of a saddle-point; motion
acquires typical near-separatrix character of relatively short velocity pulses
separated by long stays in the saddle-point vicinity. 
\item \( \Lambda _{c_{2}}>\Lambda >\Lambda _{c_{3}} \) The potential has 4 zeros
and two different types of periodic motion exist (see Fig.~\ref{figK>0k<05_2}).
Their period diverges as 
\begin{equation}
\label{pest2}
T(\Lambda )\sim 1/4\ln |\Lambda -\Lambda _{c_{2}}|\: ,
\end{equation}
 when \( \Lambda  \) approaches \( \Lambda _{c_{2}} \), Fig.~\ref{figK>0k<05period}. 
\item \( \Lambda _{c_{3}}>\Lambda  \) The potential has 4 zeros and two different
types of periodic motion exist (see Fig.~\ref{figK>0k<05_3}). As \( \Lambda \rightarrow \Lambda _{c_{3}} \)
(from both sides), the small-scale branch approaches a near-separatrix regime,
its period diverges as 
\begin{equation}
\label{pest3}
T(\Lambda )\sim \ln |\Lambda -\Lambda _{c_{3}}|\: ,
\end{equation}
for \( \Lambda =\Lambda _{c_{3}} \) this branch becomes aperiodic. 
\end{enumerate}
For the case \( k>1/2 \) only the critical value of \( X_{c_{3}} \) lies in
the physical region, and only one regime of periodic motion exists (see Fig.~\ref{figK>0k>05}).

\subsection{The case \protect\protect\protect\( K<0\protect \protect \protect \), \protect\protect\( k\neq 1/2\protect \protect \)
\label{K<0para}}

\begin{figure}[!h]
{\par\centering \resizebox*{8.5cm}{!}{\includegraphics{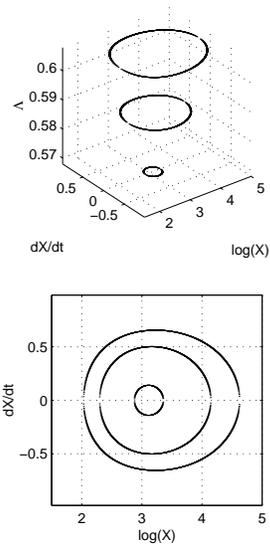}} \par}

\caption{Phase space portrait close to \protect\( \Lambda _{c_{1}}\protect \) for different
values of \protect\( \Lambda \protect \) in the case \protect\( K=-1\protect \)
and \protect\( k=0.55\protect \). The bottom figure is the top view of the
upper one. Once the first critical situation is reached a motion is possible.
\label{figK<0k>05_1}}
\end{figure}

\begin{figure}[!h]
{\par\centering \resizebox*{8.5cm}{!}{\includegraphics{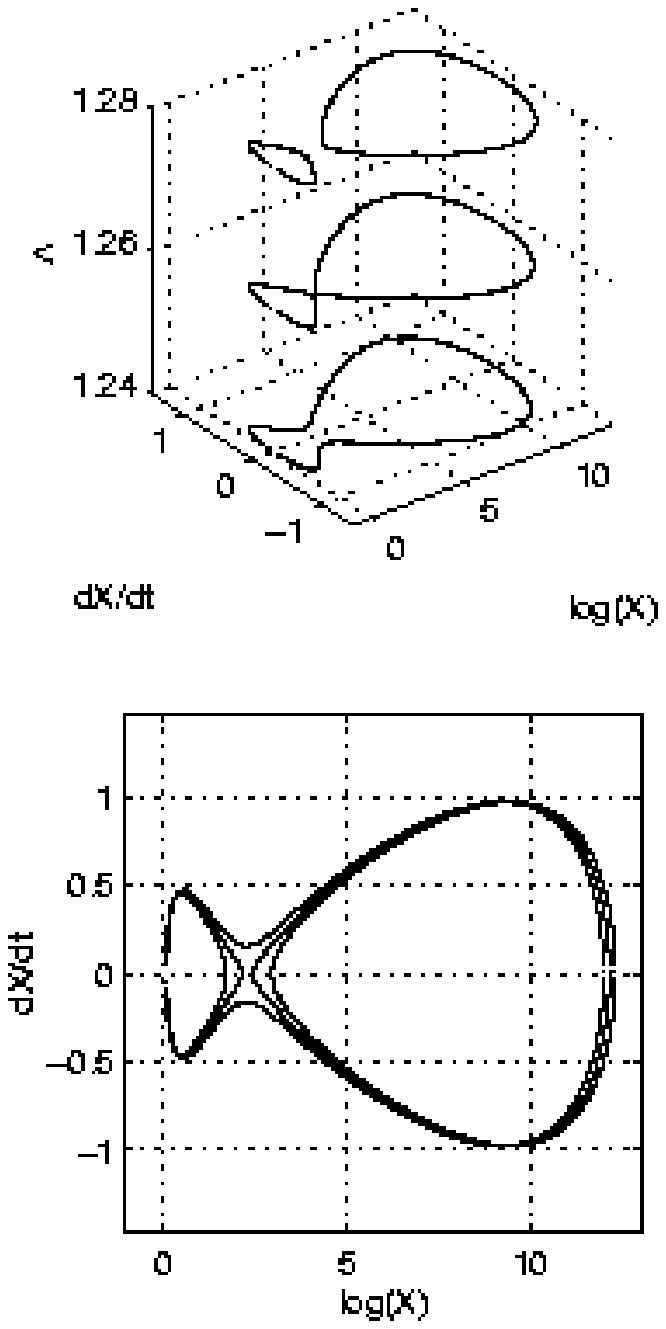}} \par}

\caption{Phase space portrait close to \protect\( \Lambda _{c_{2}}\protect \) for different
values of \protect\( \Lambda \protect \) in the case \protect\( K=-1\protect \)
and \protect\( k=0.55\protect \). The bottom figure is the top view of the
upper one. Once the second critical situation is reached, the single possible
trajectory reaches the separatrix, and the splitting in two possible motions
occurs. \label{figK<0k>05_2}}
\end{figure}

\begin{figure}[!h]
{\par\centering \resizebox*{8.5cm}{!}{\includegraphics{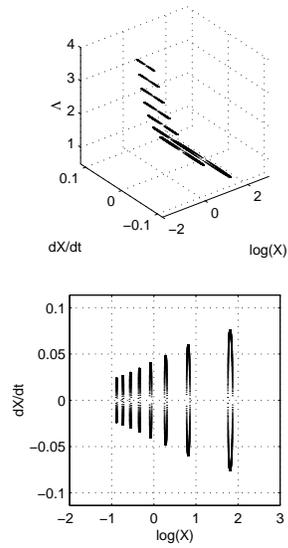}} \par}

\caption{Phase space portrait for different values of \protect\( \Lambda \protect \)
in the case \protect\( K=-1\protect \) and \protect\( k=0.2\protect \). The
bottom figure is the top view of the upper one. \label{figK<0k<05} The value
of \protect\( \Lambda \protect \) determines the typical length scale of the
periodic motion (the value of \protect\( K\protect \) is fixed).}
\end{figure}

When the total vortex angular momentum is negative, the critical root (\ref{criticsitu3})
lies outside the physical region, and therefore we expect less variety of motion
types. Otherwise, this case is similar to the \( K>0 \) case, in the sense
that most phenomena described for \( K>0 \) occur, but in a reverse order;
for instance, when \( k>1/2 \) we have three different regimes, which are analogue
to the regimes of \( K>0 \), \( k<1/2 \) case. The different regimes, illustrated
by their phase portraits, are listed below (See Table \ref{Table1}):

\begin{enumerate}
\item \( \Lambda <\Lambda _{c_{1}} \) Motion impossible. 
\item \( \Lambda _{c_{2}}<\Lambda <\Lambda _{c_{1}} \) The potential has 2 zeros
and a periodic motion is possible (see Fig.~\ref{figK<0k>05_1}). In vicinity
of a saddle point (\( \Lambda \approx \Lambda _{c_{2}} \)) motion period diverges
in the same way as in \( K>0 \) case. 
\item \( \Lambda _{c_{2}}<\Lambda  \) The potential has 4 zeros and two different
periodic motions are possible (see Fig.~\ref{figK<0k>05_2}). 
\end{enumerate}
On the other hand the situation when \( k<1/2 \), is analogous to the \( K>0 \)
and \( k>1/2 \) situation. Namely, non of the critical values are physical,
and we have only one type of periodic motion for all range of energies, see
Fig.~\ref{figK<0k<05}.

Now all the possible generic situations being described, we will proceed to
the important singular cases \( K=0 \) or \( k=1/2 \).

\subsection{The case \protect\protect\protect\( K=0\protect \protect \protect \).\label{K=3D3D3D0para}}

In this situation any rescaling of length does not affect the value of \( K \),
and it is only the value of \( \Lambda  \) that controls motion scales in the
system. The expression for the effective potential \( V(X) \) can be considerably
simplified. It is convenient to introduce new variable 
\begin{equation}
\label{Uvariable}
U=\frac{X^{2}}{4k^{2}Y},
\end{equation}
 then (\ref{potential}) becomes 
\begin{equation}
\label{potforK=0}
V(U)=\frac{2}{\pi ^{2}}k^{2}\left( 1-k\right) ^{2}\left( U-\frac{1}{\left( 1-k\right) ^{2}}\right) \left( U-1\right) \: .
\end{equation}
 The zeros are simple and the different conditions imposed on \( X \) imply
a motion confined between 
\begin{equation}
\label{rangeforK=0}
1\leq U\leq \frac{1}{\left( 1-k\right) ^{2}}\: ,
\end{equation}
 In this situation double zero do not occur. Note that for exact collapse case,
\( k=1/2 \), transformation to new variable (\ref{Uvariable}) is singular
and does not work. In this case the potential \( V(X) \) is constant; its value
is negative only in a range of energies \( \Lambda \in [1/2,1] \), which means,
that collapse configurations have their energies confined to this interval.
We will return to the discussion of collapse configurations in Section 4.

\begin{figure}[!h]
{\par\centering \resizebox*{8.5cm}{!}{\includegraphics{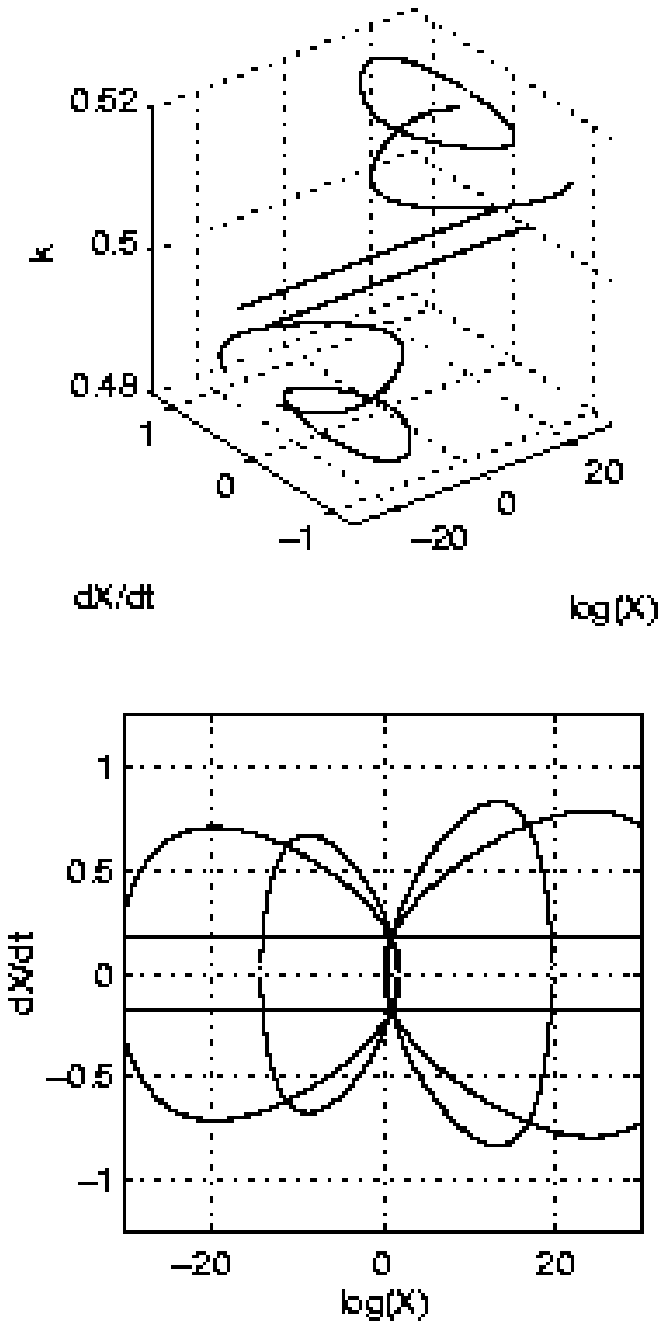}} \par}

\caption{Phase space portrait close to \protect\( \Lambda _{c_{2}}\protect \) for different
values of \protect\( k\protect \) in the case \protect\( K=0\protect \). The
bottom figure is the top view of the upper one.\label{figK_0h_hc1}For this
singular case we notice the two lines at \protect\( k=1/2\protect \) corresponding
one to collapse and one to infinite expansion. On the projection plot (bottom),
the trajectories \protect\( k<1/2\protect \) and \protect\( k>1/2\protect \)
seem to intersect on those two lines. For \protect\( \Lambda <\Lambda _{c_{2}}\protect \)
collapse is not possible, and the trajectories do not intersect. We notice that
close to the collapse condition \protect\( k=1/2\protect \), the values \protect\( k<1/2\protect \)
describe a collapse motion, while \protect\( k>1/2\protect \) describes the
expansion motion.}
\end{figure}

\begin{figure}[!h]
{\par\centering \resizebox*{8.5cm}{!}{\includegraphics{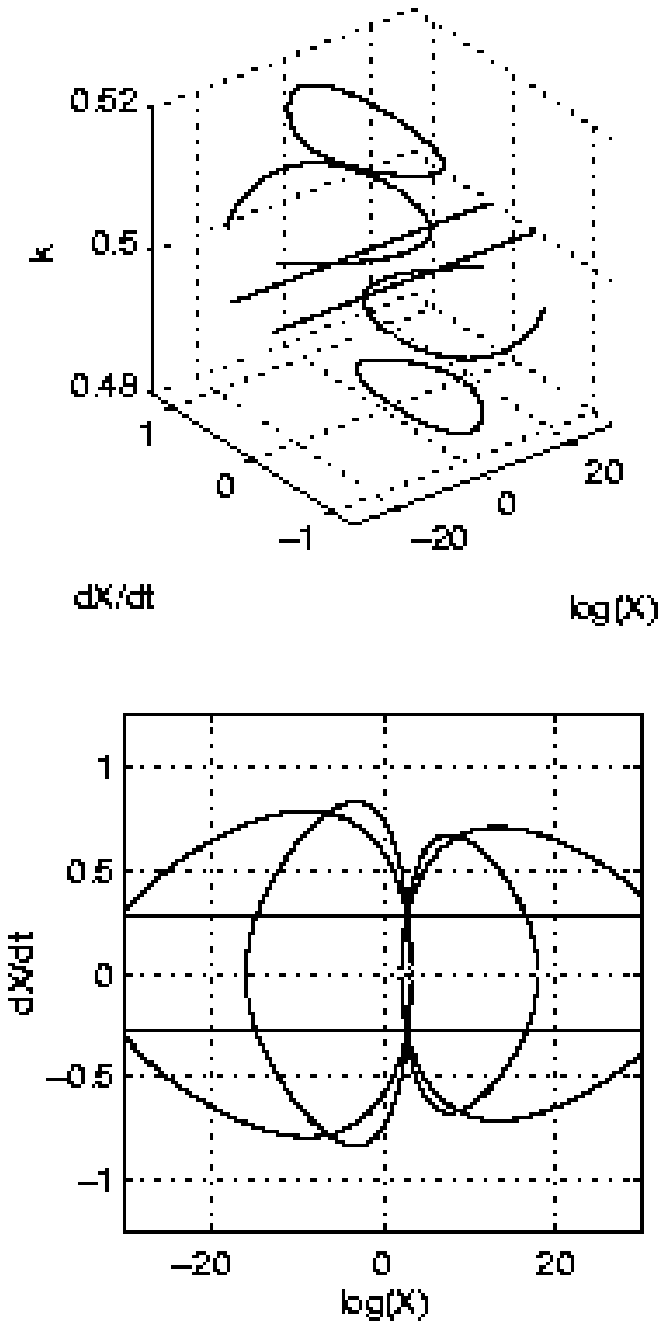}} \par}

\caption{Phase space portrait close to \protect\( \Lambda _{c_{1}}\protect \) for different
values of \protect\( k\protect \) for the case \protect\( K=0\protect \).
The bottom figure is the top view of the upper one.\label{FigK=3D0h=3Dhc2}For
this singular case we notice the two lines at \protect\( k=1/2\protect \) corresponding
one to collapse and one to infinite expansion. On the projection plot (bottom),
the trajectories \protect\( k<1/2\protect \) and \protect\( k>1/2\protect \)
seem to intersect on those two lines. For \protect\( \Lambda >\Lambda _{c_{1}}\protect \)
collapse is not possible, and the trajectories do not intersect. We notice that
close to the collapse condition \protect\( k=1/2\protect \), the values \protect\( k<1/2\protect \)
describe a collapse motion, while \protect\( k>1/2\protect \) describes the
expansion motion.}
\end{figure}

In this singular case, to visualize the approach to collapse (\( k\rightarrow 1/2 \)),
it is better to have \( k \) fluctuating and \( \Lambda  \) fixed. This is
due to the fact that, when \( k=1/2 \), which corresponds to the merging conditions,
the motion is possible only within a range of energies \( \Lambda _{2}>\Lambda >\Lambda _{1} \).
The phase portrait of trajectories taken for different values of \( k \) is
shown in Fig.~\ref{figK_0h_hc1} and Fig.~\ref{FigK=3D0h=3Dhc2}. These two
figures are taken close to the values \( \Lambda =1/2 \) and \( \Lambda =1 \);
we notice for the \( k=1/2 \) case two straight lines corresponding to the
two possible motions of expansion or collapse. Note, that when \( \Lambda \notin [1/2\: ,\; 1] \)
a collapse cannot occur, which means that the trajectories for \( k<1/2 \)
and \( k>1/2 \) are similar to those shown in Fig.~\ref{figK_0h_hc1}, but
do not intersect on the top view, the results are summarized in Table \ref{Table2}.

\subsection{The case \protect\protect\protect\( k=1/2\protect \protect \protect \)\label{k=3D3D3D1/2para}}

\begin{figure}[!h]
{\par\centering \resizebox*{8.5cm}{!}{\includegraphics{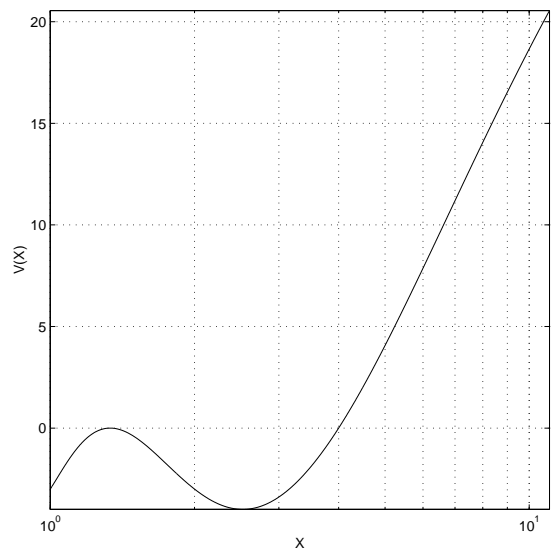}} \par}

\caption{The effective potential \protect\( V(X)\protect \) in the singular case \protect\( k=1/2\protect \)
(\protect\( \delta =0\protect \)) for the critical energy corresponding to
\protect\( \Lambda =1/4\protect \): the double root case.\label{figk=3D1/2lam=3D1/4}}
\end{figure}

\begin{figure}[!h]
{\par\centering \resizebox*{8.5cm}{!}{\includegraphics{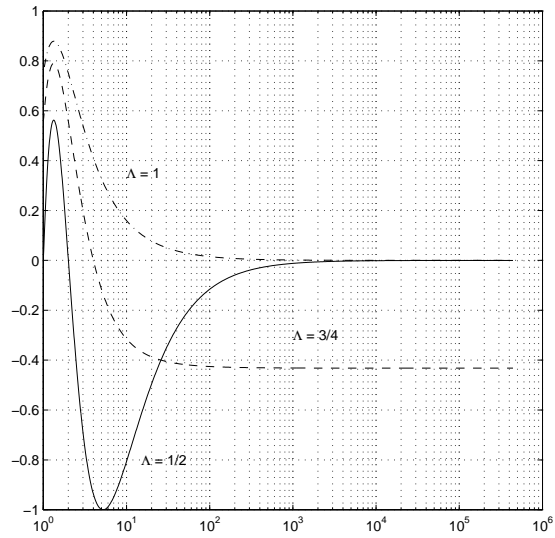}} \par}

\caption{Evolution of the effective potential \protect\( V(X)\protect \) in the singular
case \protect\( k=1/2\protect \) (\protect\( \delta =0\protect \)) within
the energy range corresponding to possible collapse \protect\( \Lambda \in [1/2,1]\protect \).
We notice that unbounded motion occurs. \label{figk=3D1/2lam=3D1/2}}
\end{figure}

\begin{figure}[!h]
{\par\centering \resizebox*{8.5cm}{!}{\includegraphics{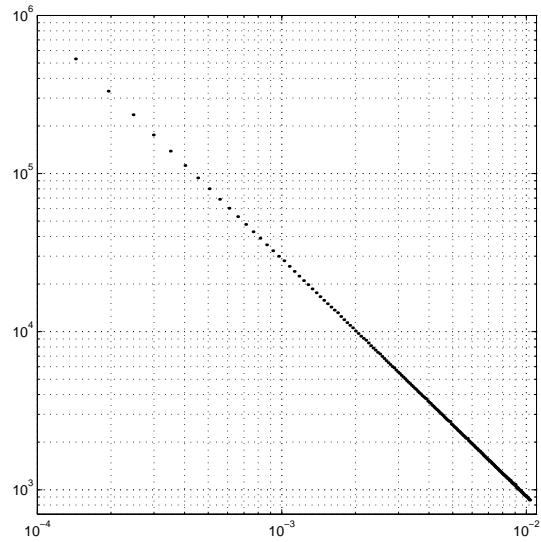}} \par}

\caption{Period of the periodic motion (\protect\( \Lambda <1/2\protect \)) motion
versus \protect\( |\Lambda -1/2|\protect \) in the singular case \protect\( k=1/2\protect \)
(\protect\( \delta =0\protect \)). We notice a power-law divergence of the
period as we approach collapse. The measured exponent is \protect\( 3/2\protect \).
See appendix \ref{appendpowerlawdiv} for the analytical computation.\label{figperiovslamk=3D1/2}}
\end{figure}

\begin{figure}[!h]
{\par\centering \resizebox*{8.5cm}{!}{\includegraphics{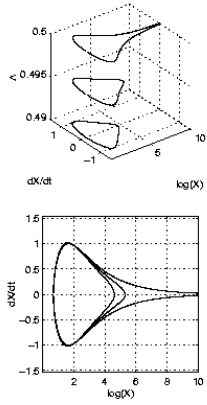}} \par}

\caption{Phase space portrait for different values of \protect\( \Lambda \protect \)
close to \protect\( \Lambda _{c}=1/2\protect \), in the case \protect\( K=1\protect \)
and \protect\( k=1/2\protect \) (\protect\( \delta =0\protect \)). The bottom
figure is the top view of the upper one. We notice the singular evolution of
the motion to larger scales as \protect\( \Lambda \protect \) approaches \protect\( \Lambda _{c}\protect \).\label{fig_period_k=3D1/2K>0l1/2}}
\end{figure}

This special case corresponds to ``scale invariance'', meaning that any length
can be rescaled without changing the energy of the system. Therefore the energies
of the critical situations should not depend on the value of \( K \), which
is scale dependent. The effective potential is now simply a fraction of polynomial
and can be rewritten as 
\begin{equation}
\label{potfork1/2}
V(X)=\lambda \left( X-X_{1}\right) \left( X-X_{2}\right) \left( X-X_{3}\right) \left( X-X_{4}\right) /X^{4}\: ,
\end{equation}
 where 
\begin{equation}
\label{rootsfork1/2}
X_{1}=\frac{K}{1/2-\Lambda }\: ,\; X_{2}=\frac{K}{1/2+\Lambda }\: ,\; X_{3}=\frac{K}{1-\Lambda }\: ,\; X_{4}=\frac{K}{1+\Lambda }\: ,
\end{equation}
 and 
\begin{equation}
\label{normofpot}
\lambda =\frac{(1/2+\Lambda )(1/2-\Lambda )(1+\Lambda )(1-\Lambda )}{4\pi ^{2}\Lambda ^{2}}\: .
\end{equation}
 We have three critical values, the first two \( \Lambda _{c_{1}}=1 \) and
\( \Lambda _{c_{2}}=1/2 \) correspond to the divergence of one root, while
the third \( \Lambda _{c_{3}}=1/4 \) corresponds to a double root situation
\( X_{2}=X_{3} \). Depending on the sign of \( K \) we have the following
allowed supports (See Table \ref{Table3}):

\begin{enumerate}
\item \( K>0 \) and \( \Lambda <\Lambda _{c_{3}} \) implies \( X_{1}>X>X_{2} \).
The potential \( V(X) \) for the critical situation \( \Lambda =1/4 \) is
illustrated on Fig.~\ref{figk=3D1/2lam=3D1/4}. As we approach collapse (\( K\rightarrow 0 \))
this motion vanishes. 
\item \( K>0 \) and \( \Lambda _{c_{3}}<\Lambda <\Lambda _{c_{2}} \) implies \( X_{1}>X>X_{3} \).
The potential \( V(X) \) for the critical situation \( \Lambda =1/2 \) is
illustrated on Fig.~\ref{figk=3D1/2lam=3D1/2}. As we approach collapse (\( K\rightarrow 0 \))
this motion also vanishes. Nevertheless this case is interesting as an aperiodic
motion very close to collapse is reached as \( \Lambda \rightarrow 1/2 \).
The period of the motion as a function of \( |\Lambda -1/2| \) is shown in
Fig.~\ref{figperiovslamk=3D1/2}. It diverges according to a power law: 
\begin{equation}
\label{power32}
T\sim \left( \Lambda -1/2\right) ^{-3/2}\: ,
\end{equation}
 (see the Appendix for its derivation). Phase space portraits of the trajectories
are illustrated in Fig.~\ref{fig_period_k=3D1/2K>0l1/2}. 
\item \( K>0 \) and \( \Lambda _{c_{2}}<\Lambda <\Lambda _{c_{1}} \), then \( X>X_{3} \).
This motion is the closest to the collapse, as infinite expansion is possible,
but \( X \) is bounded from below. In this situation we may see a collapse
course, which rebounds on \( X_{3} \) and goes for an infinite expansion. As
we approach collapse (\( K\rightarrow 0 \)) \( X_{3}\rightarrow 0 \), and
depending on the sign of \( \dot{X} \) a full collapse can occur. 
\item \( K>0 \), \( \Lambda _{c_{1}}<\Lambda  \) and \( K<0 \), \( \Lambda <\Lambda _{c_{2}} \)
motion impossible. 
\item \( K<0 \) and \( \Lambda _{c_{2}}<\Lambda <\Lambda _{c_{1}} \) , then \( X>X_{1} \).
We have a similar situation as the one discussed for the \( K>0 \) and \( \Lambda _{c_{2}}<\Lambda <\Lambda _{c_{1}} \). 
\item \( K<0 \) and \( \Lambda _{c_{1}}<\Lambda  \), then \( X_{3}>X>X_{1} \).
Here we can anticipate a similar situation as the one discussed for the \( K>0 \)
and \( \Lambda _{c_{3}}<\Lambda <\Lambda _{c_{2}} \). 
\end{enumerate}
In this section all possible types of motion have been discussed, all the results
for the different cases are summarized in the Tables \ref{Table1}-\ref{Table3}.
A more detailed discussion of the vortices behavior and trajectories, as the
collapse configuration is approached, is presented in the next Section. 

\newpage

\begin{table}[!h]
{\centering \begin{tabular}{|c|c|c|c|c|}
\hline 
\( \Lambda  \)&
Type of motion&
Period&
\( K \)&
\( k \)\\
\hline 
\hline 
\( \Lambda >\Lambda _{c_{1}} \)&
no motion&
no period&
\( K>0 \)&
\( k<1/2 \)\\
\hline 
\( \Lambda _{c_{1}}>\Lambda >\Lambda _{c_{2}} \)&
periodic motion&
\( T\approx 1/2\ln |\Lambda -\Lambda _{c_{2}}| \)&
\( K>0 \)&
\( k<1/2 \)\\
\hline 
\( \Lambda _{c_{2}}>\Lambda >\Lambda _{c_{3}} \)&
2 different periodic motions, &
\( T\approx 1/4\ln |\Lambda -\Lambda _{c_{2}}| \)&
\( K>0 \)&
\( k<1/2 \)\\
&
1 large scale, 1 small scale&
&
&
\\
\hline 
\( \Lambda _{c_{3}}>\Lambda  \)&
2 different periodic motions, &
\( T\approx \ln |\Lambda -\Lambda _{c_{3}}| \)&
\( K>0 \)&
\( k<1/2 \)\\
&
1 large scale, 1 small scale&
&
&
\\
\hline 
\hline 
any \( \Lambda  \)&
 periodic motion &
no singularities for \( T \)&
\( K>0 \)&
\( k>1/2 \)\\
\hline 
\hline 
\( \Lambda <\Lambda _{c_{1}} \)&
no motion&
no period&
\( K<0 \)&
\( k>1/2 \)\\
\hline 
\( \Lambda _{c_{1}}<\Lambda <\Lambda _{c_{2}} \)&
periodic motion&
\( T\approx 1/2\ln |\Lambda -\Lambda _{c_{2}}| \)&
\( K<0 \)&
\( k>1/2 \)\\
\hline 
\( \Lambda _{c_{2}}<\Lambda  \)&
2 different periodic motions, &
\( T\approx 1/4\ln |\Lambda -\Lambda _{c_{2}}| \)&
\( K<0 \)&
\( k>1/2 \)\\
&
1 large scale, 1 small scale&
&
&
\\
\hline 
\hline 
any \( \Lambda  \)&
1 periodic motion &
no singularities for \( T \)&
\( K<0 \)&
\( k<1/2 \)\\
\hline 
\end{tabular}\par}

\caption{Different types of motion for the general situation \protect\( K\ne 0\: ,k\ne 1/2\protect \).
The different critical values are }

{\par\centering \( \Lambda _{c_{1}}=(2\delta /K)^{2\delta } \), \( \Lambda _{c_{2}}=\frac{1}{2}(1+2\delta )(2\delta /K)^{2\delta } \),
\( \Lambda _{c_{3}}=\frac{1}{4}[(3+2\delta )/K]^{2\delta } \).\label{Table1}\par}
\end{table}

\begin{table}[!h]
{\centering \begin{tabular}{|c|c|c|c|c|}
\hline 
\( \Lambda  \)&
Type of motion&
Period&
\( K \)&
\( k \)\\
\hline 
\hline 
any \( \Lambda  \)&
periodic motion, small scales&
period found numerically&
\( K=0 \)&
\( k<1/2 \)\\
\hline 
\( 1/2<\Lambda <1 \)&
collapse or expansion&
aperiodic motion&
\( K=0 \)&
\( k=1/2 \)\\
\hline 
any \( \Lambda  \)&
periodic motion, large scales&
period found numerically&
\( K=0 \)&
\( k>1/2 \)\\
\hline 
\end{tabular}\par}

\caption{Different types of motion for the special situation \protect\( K=0\: ,k\ne 1/2\protect \).\label{Table2}}
\end{table}

\newpage

\begin{table}[!h]
{\centering \begin{tabular}{|c|c|c|c|c|}
\hline 
\( \Lambda  \)&
Type of motion&
Period&
\( K \)&
\( k \)\\
\hline 
\hline 
\( \Lambda <\Lambda _{c_{3}} \)&
periodic motion&
\( T\approx \ln |\Lambda -\Lambda _{c_{3}}| \)&
\( K>0 \)&
\( k=1/2 \)\\
\hline 
\( \Lambda _{c_{3}}<\Lambda <\Lambda _{c_{2}} \)&
periodic motion&
\( T\approx |\Lambda -\Lambda _{c_{2}}|^{-3/2} \)&
\( K>0 \)&
\( k=1/2 \)\\
\hline 
\( \Lambda _{c_{2}}<\Lambda <\Lambda _{c_{1}} \)&
aperiodic motion&
no period \( ^{*} \)&
\( K>0 \)&
\( k=1/2 \)\\
\hline 
\( \Lambda _{c_{1}}<\Lambda  \)&
no motion&
no period&
\( K>0 \)&
\( k=1/2 \)\\
\hline 
\hline 
\( \Lambda <\Lambda _{c_{2}} \)&
no motion&
no period&
\( K<0 \)&
\( k=1/2 \)\\
\hline 
\( \Lambda _{c_{2}}<\Lambda <\Lambda _{c_{1}} \)&
aperiodic motion&
no period \( ^{**} \)&
\( K<0 \)&
\( k=1/2 \)\\
\hline 
\( \Lambda _{c_{1}}<\Lambda  \)&
periodic motion&
\( T\approx |\Lambda -\Lambda _{c_{1}}|^{-3/2} \)&
\( K<0 \)&
\( k=1/2 \)\\
\hline 
\end{tabular}\par}

\caption{Different types of motion for the special situation \protect\( k=1/2\: ,K\ne 0\protect \). }

The different critical values are \( \Lambda _{c_{1}}=1 \), \( \Lambda _{c_{2}}=1/2 \),
\( \Lambda _{c_{3}}=1/4 \). 

\( ^{*} \) minimum approach \( X=X_{3}=K/(1-\Lambda ) \). 

\( ^{**} \) minimum approach \( X=X_{1}=K/(1/2-\Lambda ) \).\label{Table3}
\end{table}

\newpage

\section{Collapse scenarii}

As we have mentioned earlier, for the collapse to happen, two \char`\"{}resonance
conditions\char`\"{} have to be satisfied. First, the sum of the vortex strengths
inverses have to be zero. This condition does not impose any geometrical restrictions
on vortex positions, for a given system of vortices it is either satisfied or
not. Initial positions leading to collapse are specified by the second collapse
condition, \( K=0 \). It also defines the range of energies, for which the
self-similar dynamics can occur (for \( k=1/2 \)), but does not tell in which
direction (collapse or expansion) it will go.

For a given vortex configuration, let us consider a coordinate system with \( x \)-axis
passing through the two positive vortices (\( k_{1}=k_{2}=1 \)) and an origin
in the middle between them, so that positive vortices have coordinates \( (-d/2,0) \)
and \( (d/2,0) \), where \( d \) is the distance between them. Denoting the
coordinates of the third vortex by \( (x,y) \) we can rewrite the condition
\( K=0 \) as 
\begin{equation}
\label{K0circle}
x^{2}+y^{2}=\frac{d^{2}}{2k}(1-k/2)
\end{equation}
 i.e. third vortex has to lie on a \char`\"{}critical circle\char`\"{} centered
at the midpoint between the two positive ones, with a radius \( (d/2)\sqrt{2/k-1} \).

\begin{figure}[!h]
{\par\centering \resizebox*{8.5cm}{!}{\includegraphics{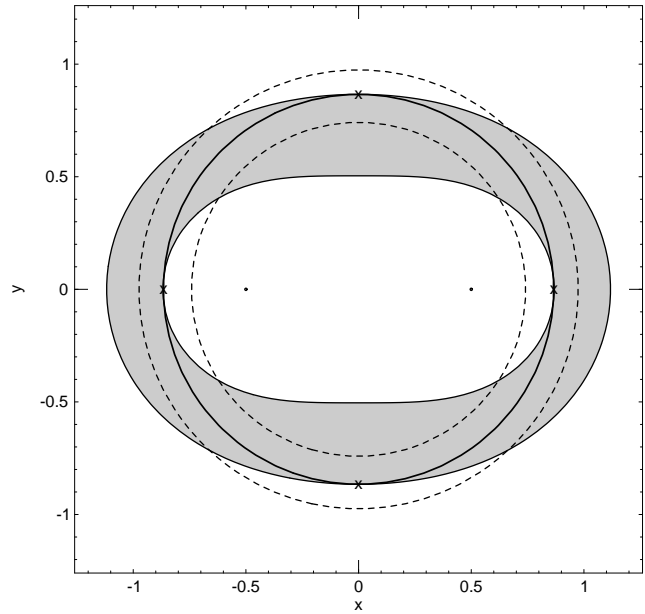}} \par}

\caption{The approach to collapse. Initial positions available for the vortex with negative
strength. The two vortex with strength \protect\( 1\protect \) are fixed, and
located by the two thick points on the plot. Circles refer to different values
of the constant \protect\( K\protect \), while the oval curves correspond to
different values of \protect\( \Lambda \protect \). The shaded region corresponds
to initial conditions leading to the aperiodic infinite expansion type of motion.\label{Initialcondforcollapse}}
\end{figure}

In the ``resonant'' case \( k=1/2 \), the critical circle represents a set
of initial conditions leading to a self-similar collapse or expansion. Level
lines of \( \Lambda  \) cross the circle in four points for \( \Lambda \in (1/2,1) \),
see Figure~\ref{Initialcondforcollapse}. These points are reflections of each
other in the coordinate axes, this symmetry lead the coordinate axis to divide
the plane in four dynamically equivalent quadrants. A reflection of a vortex
configuration is equivalent to changing the direction of time, so that points
in adjacent quadrants have opposite directions of their dynamics. From the original
equations (\ref{equmotion1}), we deduce, that the parts of the critical circle
lying in the I and III quadrants lead to a finite-time collapse, and those in
II and IV quadrants to an infinite self-similar expansion.

Four intersections of the critical circle with the coordinate axes are equilibrium
positions, where \( \Lambda  \) reaches its limiting values (on the circle);
the maximum value \( \Lambda =1 \) corresponds to the equilateral triangle,
and the minimum \( \Lambda =1/2 \) to collinear configuration.

\begin{figure}[!h]
{\par\centering \resizebox*{8.5cm}{!}{\includegraphics{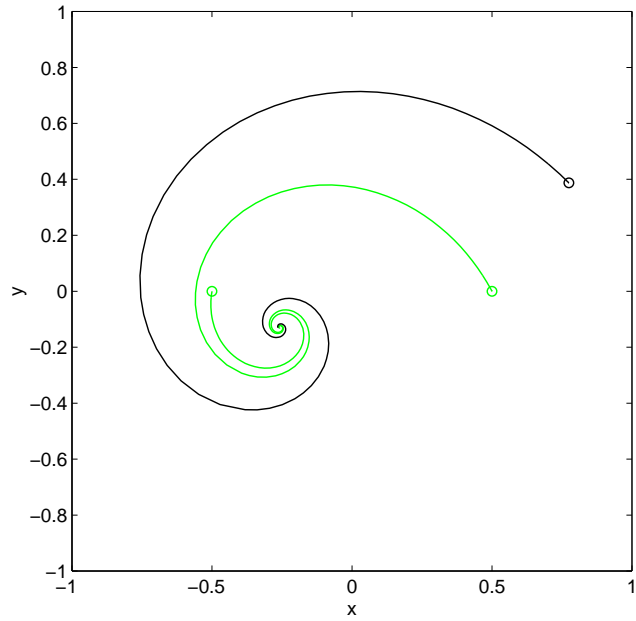}} \par}

\caption{Vortex trajectories during collapse. Initial positions are shown by circles,
they correspond to a fastest collapse value \protect\( \Lambda =\sqrt{3}/2\protect \);
negative vortex trajectory is plotted in black, positive ones in gray. All three
vortices converge to the center of vorticity in finite time \protect\( \tau =4\pi /3\protect \).\label{figA1} }
\end{figure}

We define a rate of collapse as a rate of change of the squared distance between
the two positive vortices, i.e. as \( \dot{X} \). In this case, self-similar
motion will have constant rate of collapse. Indeed, vortex velocities are inversely
proportional to the inter-vortex distances, so that \( \dot{X}=2R_{1}\dot{R}_{1} \)
is independent of the scale of the motion. The rate of collapse is determined
by the vortex energy: 
\begin{equation}
\label{rateofcollapse}
\dot{X}(\Lambda )=\sqrt{4(1-\Lambda ^{2})(1/4-\Lambda ^{2})},
\end{equation}
 it tends to zero when \( \Lambda  \) approaches one of the equilibrium values
\( \Lambda =1/2,\: 1 \). Figure~\ref{figA1} shows vortex trajectories for
the fastest collapse case \( \Lambda =\sqrt{3}/2 \).

\begin{figure}[!h]
{\par\centering \resizebox*{8.5cm}{!}{\includegraphics{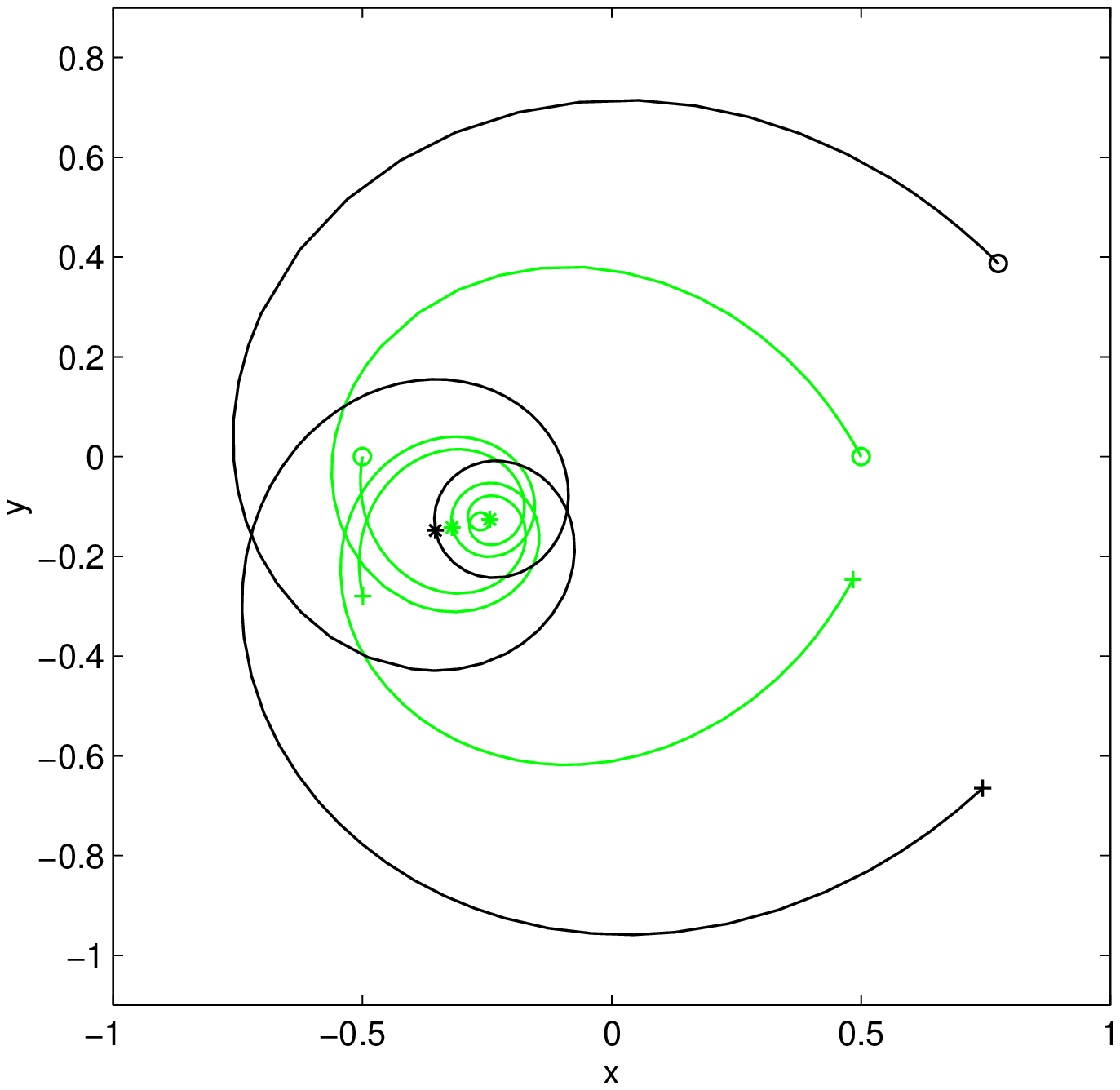}} \par}

{\par\centering \resizebox*{8.5cm}{!}{\includegraphics{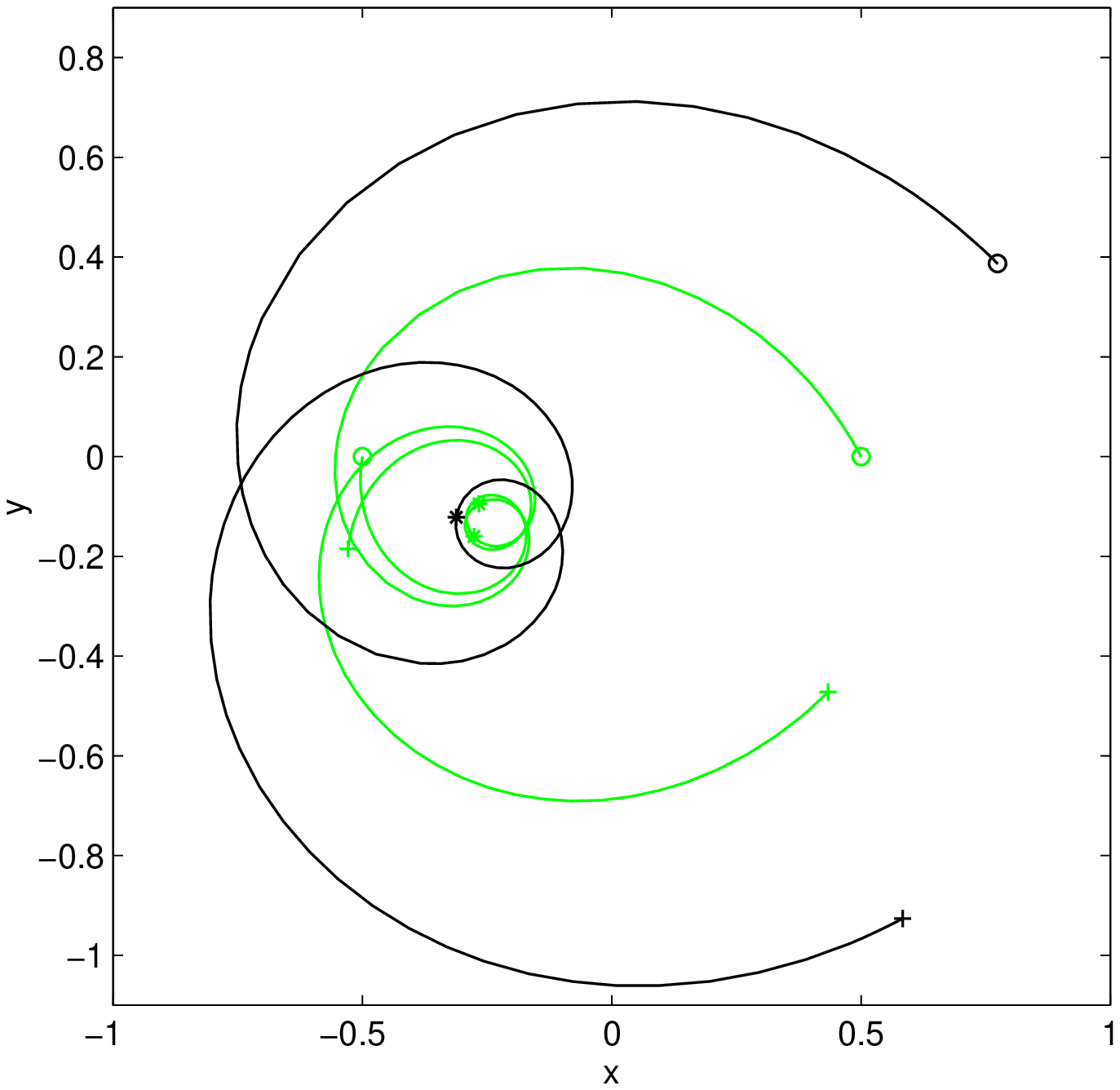}} \par}

\caption{Unbounded near-collapse motion for \protect\( k=1/2\protect \). Negative vortex:
black; positive: gray. Closest approach positions are marked by asterisks.}

Upper plot \( K<0 \): vortices contract until a collinear configuration with
$X_{min}=K/(1/2-\Lambda)$ is reached.

Lower plot \( K>0 \): closest approach is this case is $X_{min}=K/(1-\Lambda)$ and
corresponds to an isosceles triangle. \label{figureA2}
\end{figure}

When the negative vortex is slightly away from the critical circle, motion type
depends on whether \( \Lambda  \) lies in the interval \( (1/2,1) \) or not.
If \( \Lambda \in (1/2,1) \), vortices escape to infinity, their motion is
aperiodic and unbounded. If the negative vortex lies in the I or III quadrant,
vortex triangle starts to contract in a nearly self-similar way, resembling
an exact collapse trajectory, but by the time when \( X \) reaches its minimum
value (\( K/(1/2-\Lambda ) \) for \( K<0 \) or \( K/(1-\Lambda ) \) for \( K>0 \))
the triangle evolves into a collinear (\( K<0 \)) or isosceles (\( K>0 \))
configuration, as shown in Figure~\ref{figureA2}. After that vortices start
to expand ad infinitum, asymptotically approaching a self-similar expanding
trajectory. It follows, that a contracting self-similar motion in unstable,
while an expanding one is asymptotically stable, a result obtained in \cite{Tavantzis88}
via linear analysis. When \( \Lambda \notin [1/2,1] \) motion becomes periodic,
\( X \) oscillates in an interval \( (K/(1-\Lambda );K/(1/2-\Lambda )) \)
for \( K<0 \) or \( (K/(1/2-\Lambda );K/(1-\Lambda )) \) for \( K>0 \).

\begin{figure}[!h]
{\par\centering \resizebox*{8.5cm}{!}{\includegraphics{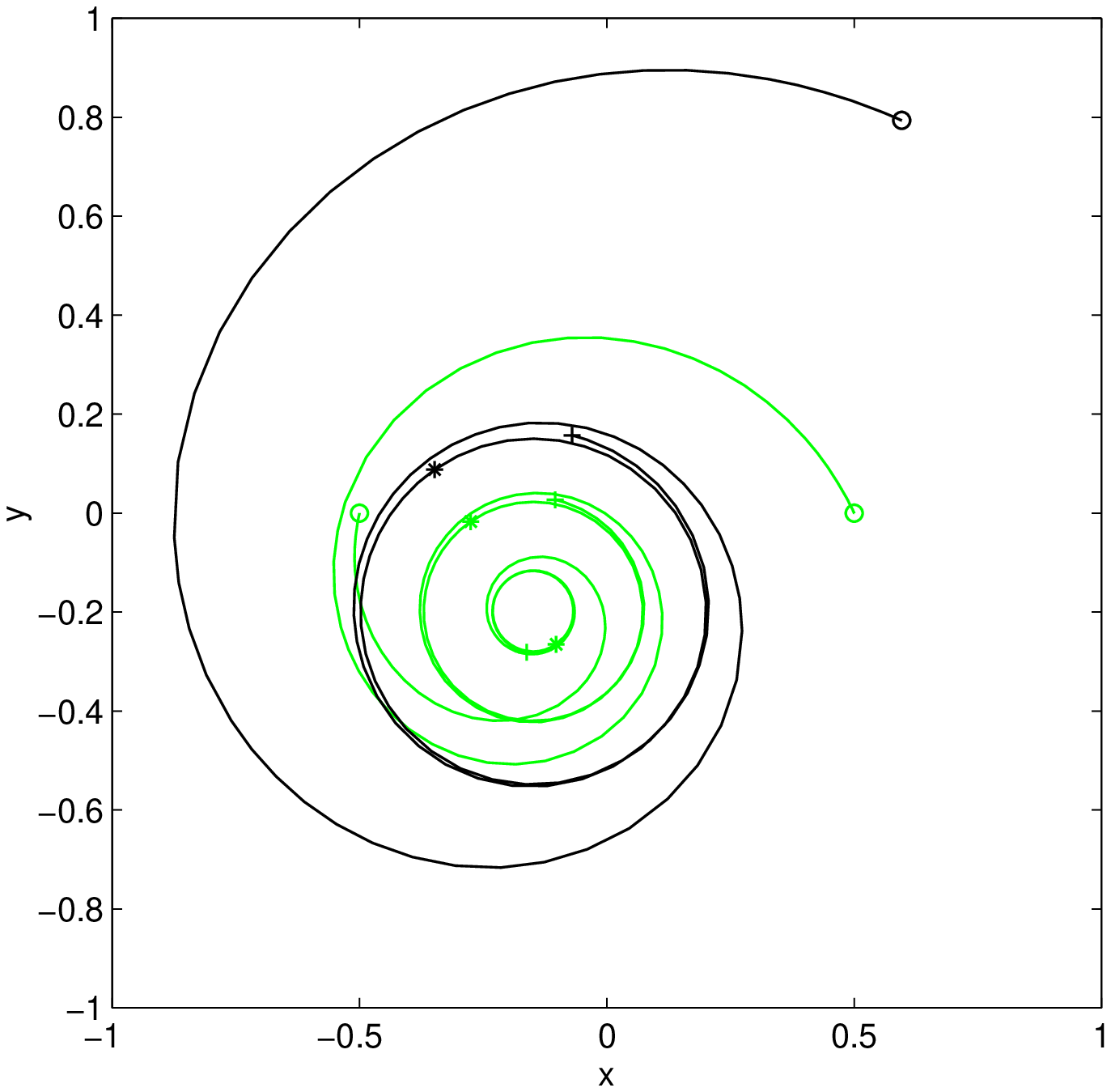}} \par}

{\par\centering \resizebox*{8.5cm}{!}{\includegraphics{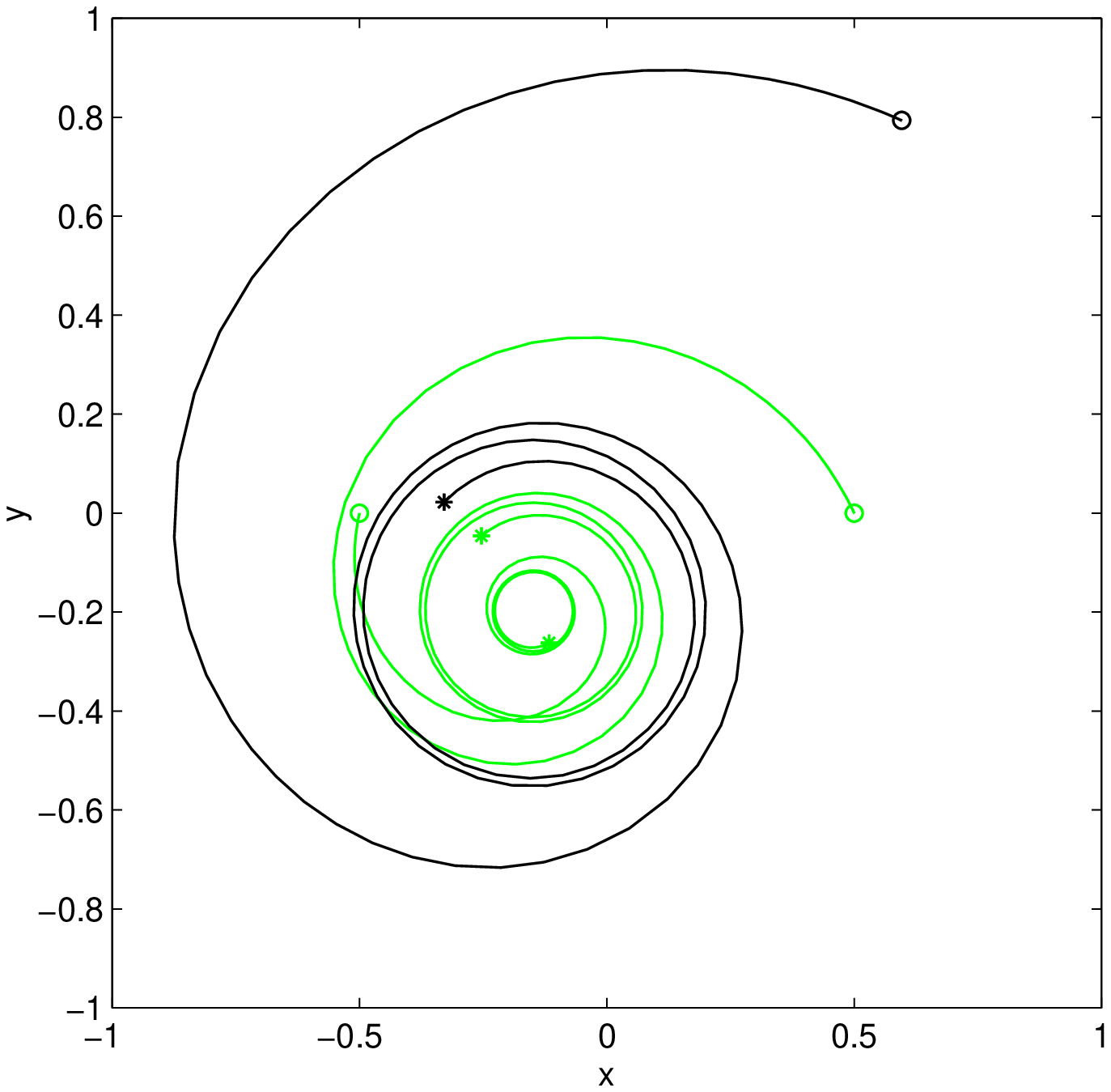}} \par}

\caption{Motion in the vicinity of a saddle-point \protect\( \Lambda =\Lambda _{c_{2}}\protect \)
Contracting vortices approach an unstable equilibrium configuration, their spiral
trajectories tighten and become nearly circular. }

Upper plot \( \Lambda <\Lambda _{c_{2}} \): vortex configuration cannot pass
through the saddle-point, and reflects back after reaching collinear configuration. 

Lower plot \( \Lambda >\Lambda _{c_{2}} \): vortices ``above'' the saddle-point,
and continue contraction (only a small piece of this part of the trajectory
is shown).\label{figureA3}
\end{figure}

In case \( k\neq 1/2 \) unbounded motion does not exist anymore, yet the ratio
of maximum vortex separation during the motion to the closest approach distance
can be arbitrarily large as \( k\rightarrow 1/2 \). An interesting phenomenon
is an existence of bounded aperiodic motions for particular values of vortex
energy. They occur when an unstable equilibrium collinear configuration \( X=X_{c_{2}} \)
appears at \( \Lambda =\Lambda _{c_{2}} \). Apart from the logarithmic divergence
of the motion period near \( \Lambda _{c_{2}} \) it also leads to a discontinuous
dependence of the closest approach distance \( X_{min} \) on initial vortex
positions. Indeed, contracting vortices with \( \Lambda =\Lambda _{c_{2}}-\epsilon  \)
where \( \epsilon  \) in an arbitrarily small positive number, will not pass
through \( X=X_{c_{2}} \), implying \( X_{min}\approx X_{c_{2}} \), while
infinitesimally different initial configuration with \( \Lambda =\Lambda _{c_{2}}+\epsilon  \)
will, leading to further contraction. This is illustrated in Figure~\ref{figureA3}.

To resume, approaching collapse means taking the limits \( K\rightarrow 0 \)
and \( k\rightarrow 0.5 \). These limits are not commutative, and for certain
situations, the result depends on how the limits are taken, e.g. these limits
may induce the divergence of \( X_{c_{1,2}} \) or set them to zero or any value.
Therefore a three vortex system can approach collapse in a number of different
ways. For instance, if we take one limit after another, we can obtain the properties
of the motion from the cases listed in Section \ref{K=3D3D3D0para} and Section \ref{k=3D3D3D1/2para}.
It is also interesting to study some of the scenarios, when the limits \( K\rightarrow 0^{+-} \)
and \( k\rightarrow 0.5^{+-} \) are taken simultaneously, leading to an interplay
of different situations, described in the previous Section. As an example, consider
a limit \( K\rightarrow 0^{+} \), \( k\rightarrow 0.5^{-} \) taken along the
line \( K/\delta =C \), where \( C \) is a constant. In this case, the saddle-point
critical values \( (X_{c_{2}},\Lambda _{c_{2}} \) tend to \( (C,1/2) \), i.e.
the scale of the inner potential well (bounded by \( X_{c_{2}} \)) stay the
same. At the same time the outer potential well spreads two infinity, and we
have a situation, when two types of motion, corresponding to the same value
of vortex energy, have entirely different length scales.

\section{Tracer advection near collapse}

\begin{figure}[!h]
{\par\centering \resizebox*{8.5cm}{!}{\includegraphics{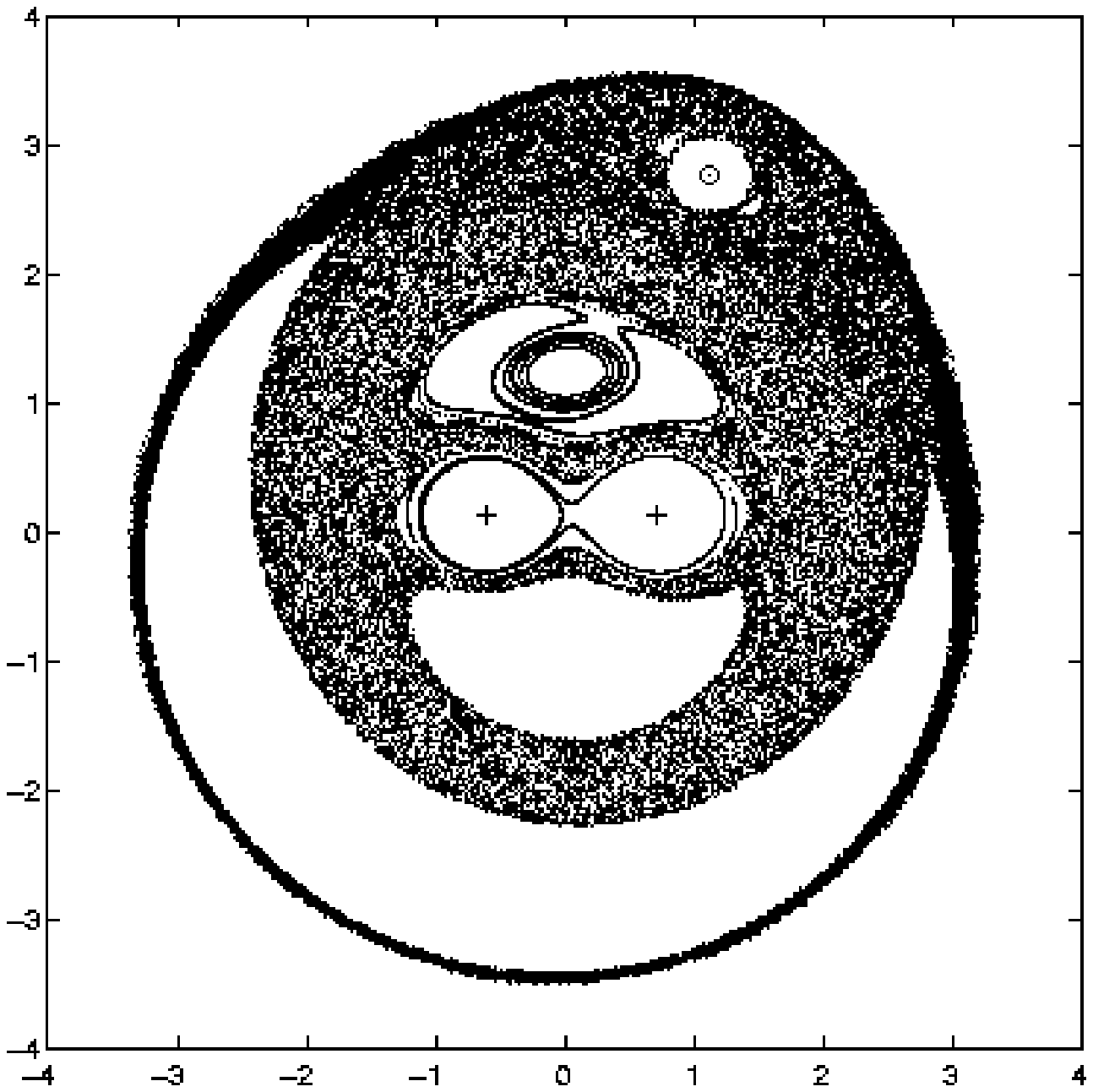}} \par}

\caption{Far from collapse \protect\( \delta =0.4\protect \). Poincaré section of \protect\( 253\protect \)
passive tracers in the flow field generated by three point vortex. The run is
over \protect\( 4000\protect \) periods. The constant of motions are \protect\( \Lambda =0.9\protect \),
\protect\( K=0\protect \). The vortex strengths are \protect\( (-0.1,\; 1,\; 1)\protect \).
The period of the motion is \protect\( T=6.59\protect \).\label{poincarek01}}
\end{figure}

\begin{figure}[!h]
{\par\centering \resizebox*{8.5cm}{!}{\includegraphics{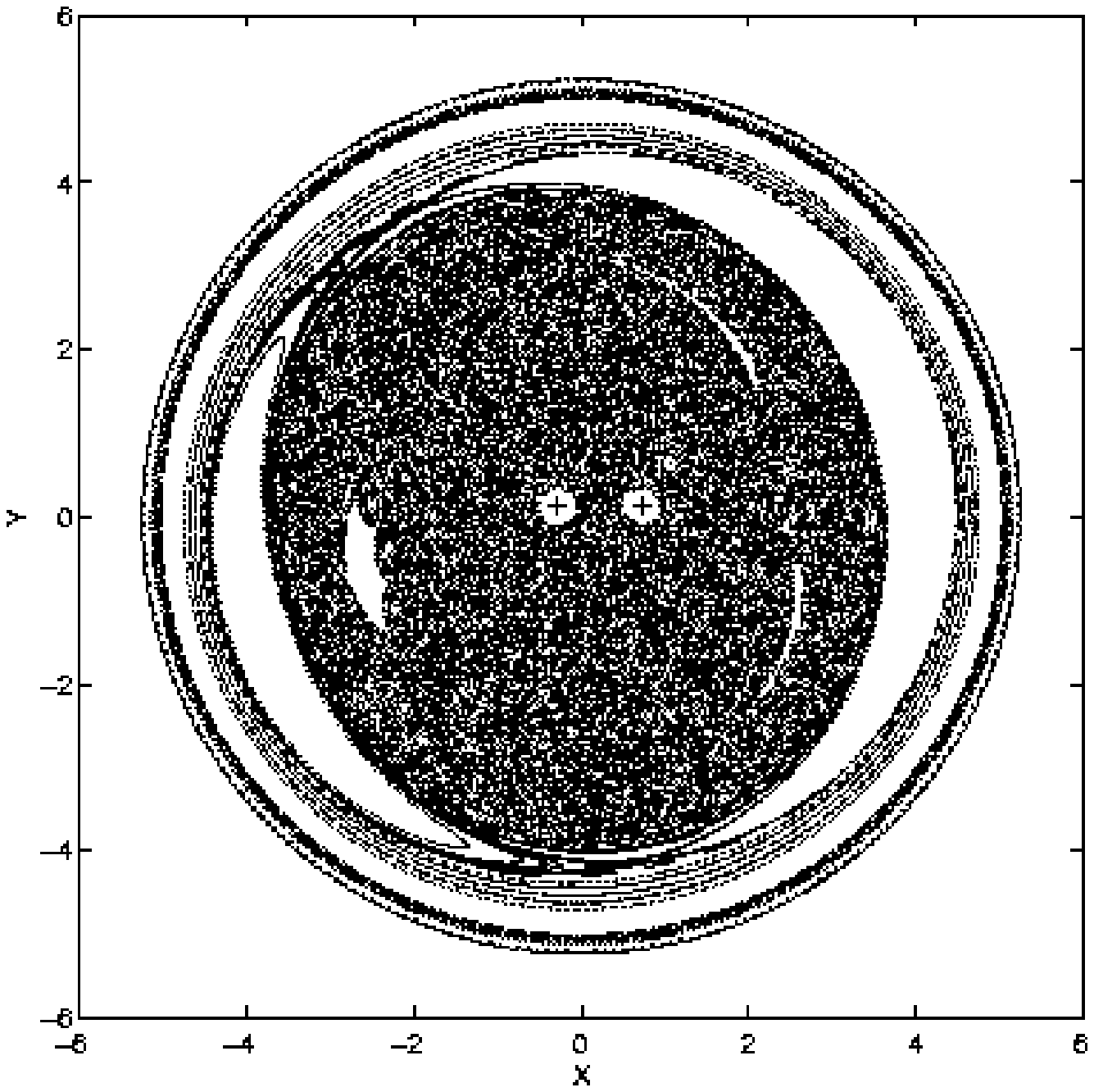}} \par}

\caption{Far from collapse \protect\( \delta =0.09\protect \). Poincaré section of
\protect\( 253\protect \) passive tracers in the flow field generated by three
point vortex. The run is over \protect\( 4000\protect \) periods. The constant
of motions are \protect\( \Lambda =0.9\protect \), \protect\( K=0\protect \).
The vortex strengths are \protect\( (-0.41,\; 1,\; 1)\protect \). The period
of the motion is \protect\( T=36.86\protect \).\label{poincarek041}}
\end{figure}

As we have mentioned, a prominent feature of near collapse dynamics is a generation
of new length scales, which can differ from the length scale of the original
configuration by orders of magnitude (see for instance (\ref{power32})). This
property has a strong influence on the mixing properties of the vortex flow.
In this Section we display some preliminary results on the difference of the
advection of passive particles (tracers) induced by collapse for a three-vortex
system. We compare the motion of passive particles (tracers) in a velocity field
of a near collapse configuration, with that in a typical ``far from collapse''
flow. A stream function of the flow, induced by moving vortices can be written
as: 
\begin{equation}
\label{streamfunc}
\psi (\textbf {x},t)=-\frac{1}{2\pi }\sum ^{3}_{i=1}\ln |\textbf {x}-\textbf {x}_{i}(t)|
\end{equation}
 corresponding equations of tracer motion are: 
\begin{equation}
\label{tracereq}
\frac{d{\textbf {r}}(t)}{dt}={\textbf {e}}_{z}\wedge \nabla \psi 
\end{equation}
 where \( \textbf {r}(t) \) denotes tracer position vector. Vortex trajectories
\( \textbf {x}_{i}(t) \) are periodic in a corotating reference frame, where
(\ref{tracereq}) have a structure of a Hamiltonian system with \( 1\frac{1}{2} \)
degrees of freedom \cite{NeufeldTel, KZ98}, so in a generic three-vortex flow
some of the tracer trajectories are regular, and some are chaotic. To visualize
advection patterns, we numerically construct Poincaré section of tracer trajectories
(in a corotating frame), which is an expedient tool for estimation of the degree
of chaotization of advection, location of the main structures in the chaotic
regions, etc. In Fig.~\ref{poincarek01}, an example of an advection pattern
of a far from collapse flow is presented. A large connected mixing region, regular
cores around vortices, elliptic islands are typical elements of a tracer phase
space in three-vortex flows \cite{NeufeldTel, KZ98}. Two most important length
scales in the advection pattern, associated with its robust features, are the
outer radius of the mixing region and the radii of vortex cores. Generally,
these two scales are comparable (their ratio is about one order of magnitude
or less). However in advection patterns of near collapse vortex configurations,
these length scales become considerably different, as the core radii tend to
zero when approaching collapse see Fig.~\ref{poincarek041}. The figures Fig.~\ref{poincarek01}
and Fig.~\ref{poincarek041} are showed to emphasize that tracers chaotic dynamics
in the near collapse case, is different than the dynamics without collapse.
A more detailed analysis will be described in another article.

\section{Conclusion}

In this paper the behavior of a three point vortex system in a near collapse
state was studied. We restricted ourselves to a special case where to vortex
are identical which allowed the reduction of the system to an effective one
dimensional Hamiltonian study. We were then able to classify all typical types
of motion which could occur in a near collapse case, which qualitatively agrees
with previous classification of the motion of three vortices \cite{Aref79}\cite{Tavantzis88}
or collapse perturbations \cite{Novikov79}. On the other hand our classification
also focus on dynamical properties of the system, which are exhibited through
the representation of trajectories in the effective phase space, giving us for
example a direct insight on the period of the motion for the closed trajectories.
It is in this sense complementary of the previous general classifications.

Furthermore, the singularity of the vortex collapse is special in the sense
that near collapse dynamics shows a number of nontrivial dynamical features.
In particular, the possibility of vorticity concentration allows one to speculate
on the fact that the collapse situation exhibits some basics mechanisms which
are of importance in \( 2D \) turbulence and its models. For instance, it becomes
clear, that three-vortex collisions play an important role in the aggregation
of vorticity in \( 2D \) turbulent flows, especially at the late stages of
freely decaying turbulence, when the vorticity occupation is rather low \cite{Sire99}.
Among these three-vortex collisions, near collapse ones are the more effective
in bringing vortex together, and are thus the most important ones; this speculation
is supported by the numerical simulations \cite{Zabusky96}. From this perspective,
we give in this paper through the analysis of the effective potential, different
typical length scales (\( X_{c} \)) and time scales (critical behaviors of
the period), which shall be of importance in models attempting to describe \( 2D \)
turbulence. For instance, the intricate structure of the vortex collapse singularity
results in a very sensitive, non trivial dependence of the closest approach
in the vicinity of collapse; it turns out that this is a discontinuous function
of initial vortex positions. This is illustrated in Section 4, by the figure
\ref{figureA3}; in the same Section the configuration of the three vortices
in their closest approach has also been discussed, and a clear distinction is
made between the collinear and isoscele triangle configuration. These results
could be of importance in deriving three-vortex merging rules analogous to the
two-vortex ones used by current punctuated Hamiltonian models of \( 2D \) turbulence.

From the dynamical point of view, the study of vortex motion near the peculiar
critical situation of collapse, has also revealed itself interesting in exhibiting
a variety of behaviors. Namely, we observed for some situations different periodic
motions for which critical behavior of the growth of the period with respect
to the control parameter \( \Lambda  \) can be logarithmic, but it also can
be as well a non-generic power-law growth with exponent \( 3/2 \) (See Section
\ref{k=3D3D3D1/2para} or Table \ref{Table3}). This last behavior is particular
as in most cases, the approach of a singular critical point leads to a typical
logarithmic divergence of the period, as for the near scattering situation discussed
in \cite{Aref79}.

Another motivation was how, the near-collapse dynamics influence the properties
of passive particles. Advection in the chaotic region of the flow due to three
identical vortices is superdiffusive \cite{KZ98}; tracers statistics is non-Gaussian,
and is characterized by power law tails in the tracer probability density function,
and algebraic decay of Poincaré recurrences statistics; both are the consequences
of the presence of long regular flights in a chaotic tracer trajectory. We expect
the advection in a general three vortex system to be anomalous too, but the
question whether or not the power law exponents and self similar behavior of
the the tracer distribution observed in \cite{KZ99} are universal, remains
open at the moment. With this in mind, it seems that the ``extreme'' dynamics
encountered in near collapse dynamics, seem to be the most appropriate candidate
to check if such universality exists. In this paper we limit ourselves to a
short illustration of the advection patterns evolve as collapse is approached.
We have observed , that as the system approaches collapse, the degree of chaotization
increases: the chaotic region grows and the area of the regular island inside
it shrinks. A more detailed study of the dynamical and statistical properties
of the advection in near-collapse flows is currently under way, and will be
presented elsewhere. 

To conclude, we emphasize once again the fact, that near collapse dynamics comprises
several types of motion. Their time and length scales differ considerably, depending
on the way the collapse conditions are violated.

\appendix

\section*{Appendix: Power-law divergence\label{appendpowerlawdiv}}

From the effective Hamiltonian (\ref{Hamilteff}) the period of the motion in
the case \( k=1/2 \) and for instance \( K>0 \) and \( \Lambda <1/2 \) (see
Fig.~\ref{figperiovslamk=3D1/2}) is then simply defined by 
\begin{equation}
\label{perioddef}
T=2\int ^{X_{1}}_{X_{3}}\frac{dX}{\sqrt{-V(X)}}\: ,
\end{equation}
 where \( X_{1} \) and \( X_{3} \) are defined in (\ref{rootsfork1/2}). To
obtain the asymptotic expression of the period as \( \Lambda \rightarrow 1/2^{-} \)(where
the \( - \) sign refers that we are taking a right limit \( \Lambda <0.5 \)),
we use some simple equivalents, and write 
\begin{equation}
\label{period_{e}quiv}
T\sim 2\lambda ^{-1/2}\int _{1}^{X_{1}}\frac{dX}{\sqrt{X_{1}/X-1}}\sim 4\lambda ^{-1/2}X_{1}\: ,
\end{equation}
 using the expression (\ref{rootsfork1/2}) for \( X_{1} \) and (\ref{normofpot})
for \( \lambda  \), we obtain the asymptotic behavior of the period as a function
of \( \Lambda  \) and \( K \)
\begin{equation}
\label{periodasympt}
T\sim K\sqrt{\frac{3}{2}}\left( \frac{1}{2}-\Lambda \right) ^{-3/2}\: .
\end{equation}
 The period diverges as \( (1/2-\Lambda )^{-3/2} \) when \( \Lambda  \) approaches
the collapse value of \( 1/2 \), and \( K\ne0  \). Note that the period is
linear in \( K \).

\section*{Acknowledgments}

This work was supported by the US Department of Navy, Grant No. N00014-96-1-0055,
and the US Department of Energy, Grant No. DE-FG02-92ER54184.

\newpage

\newpage

\newpage

\newpage

\newpage

\newpage

\newpage

\newpage

\newpage

\newpage

\newpage

\newpage

\newpage

\newpage

\newpage

\newpage

\newpage

\newpage

\newpage

\newpage

\newpage

\newpage

\end{document}